\documentclass[journal]{IEEEtran}

\usepackage{ucs}
\usepackage[utf8x]{inputenc}
\usepackage[cmex10]{amsmath}
\usepackage{cite, amsfonts, amssymb, bm, bbm, graphicx, relsize, multirow, booktabs, enumitem, algorithmic, algorithm, color, soul, multirow}
\usepackage[american]{babel}
\usepackage[T1]{fontenc}

\usepackage{amsthm}
\makeatletter 
\def\@endtheorem{\qed\endtrivlist\@endpefalse } 
\makeatother

\theoremstyle{definition}

\theoremstyle{remark}
\newtheorem{remark}{Remark}
\newtheorem{example}{Example}

\newcommand*{\herm}{^{\mathsf{H}}}
\newcommand*{\transp}{^{\mathsf{T}}}

\DeclareMathOperator{\trace}{Tr}
\DeclareMathOperator*{\argmin}{\arg\min}
\DeclareMathOperator*{\argmax}{\arg\max}

\newcommand{\e}{\mathrm{e}}
\renewcommand{\i}{\mathrm{i}}

\IEEEoverridecommandlockouts

\title{Radar-Enabled Ambient Backscatter Communications}
\author{ Luca~Venturino,~\IEEEmembership{Senior~Member,~IEEE}, Emanuele~Grossi,~\IEEEmembership{Senior~Member,~IEEE}, Marco~Lops,~\IEEEmembership{Fellow,~IEEE}, Jeremy~Johnston, Xiaodong~Wang,~\IEEEmembership{Fellow,~IEEE} 
\thanks{L.~Venturino and E.~Grossi are with the Department of Electrical and Information Engineering, University of Cassino and Southern Lazio, 03043 Cassino, Italy, and with CNIT, 43124 Parma, Italy (e-mail: l.venturino@unicas.it; e.grossi@unicas.it). M.~Lops is with the Department of Electrical and Information Technology, University of Naples Federico II, 80138 Naples, Italy, and with CNIT, 43124 Parma, Italy (e-mail: lops@unina.it). J.~Johnston and X.~Wang are with the Department of Electrical Engineering, Columbia University, New York, NY 10027, United States (e-mail: j.johnston@columbia.edu; xw2008@columbia.edu). }

\thanks{The work of M.~Lops was partially supported by the European Union under the Italian National Recovery and Resilience Plan (NRRP) of NextGenerationEU, partnership on ``Telecommunications of the Future'' (PE00000001 - program ``RESTAR'' - E63C22002040007). The work of L.~Venturino and E.~Grossi was partially supported by the Italian Ministry of Education, University, and Research under the Research Program ``Dipartimenti di Eccellenza 2018-2022.''}

\thanks{Part of the results in this work have been presented at the 2022 Asilomar Conference on Signals, Systems, and Computers~\cite{Asilomar2022}.}
}

\begin{document}
\bstctlcite{BSTcontrol}		
\maketitle	
\IEEEpeerreviewmaketitle

\begin{abstract}
In this work, we exploit the radar clutter (i.e., the ensemble of echoes generated by the terrain and/or the surrounding objects in response to the signal emitted by a radar transmitter) as a carrier signal to enable an ambient backscatter communication from a source (tag) to a destination (reader). The proposed idea relies on the fact that, since the radar excitation is periodic, the radar clutter is itself periodic over time scales shorter than the coherence time of the environment. Upon deriving a convenient signal model, we propose two encoding/decoding schemes that do not require any coordination with the radar transmitter or knowledge of the radar waveform. Different tradeoffs in terms of transmission rate and error probability can be obtained upon changing the control signal driving the tag switch or the adopted encoding rule; also, multiple tags can be accommodated with either a  sourced or an unsourced multiple access strategy. Some illustrative examples are provided.
\end{abstract}


\section{Introduction}

Communications and radar have historically followed parallel paths, which intersected only occasionally, until the emerging Internet of Things (IoT) and perceptive mobile networks have tightly intertwined them~\cite{Zhang-2021,Leyva-2021,Poor-2022} to support a number of advanced applications (such as, e.g., autonomous driving, smart cities/factories, environmental/home monitoring, healthcare). Spectrum overcrowding has been the inevitable counterbalance to these developments, posing new challenges for a more efficient exploitation of the available spectrum {\em and} a containment of operational costs, power consumption, and electromagnetic emissions. 

A response to these challenges has been the paradigm shift from spectrum sharing between two autonomous systems, possibly exchanging information and interacting with each other~\cite{Lops-2019}, to integrated sensing and communications (ISAC) architectures~\cite{Hassanien-2016,Liu-2020} encompassing just one active transmitter and different receiving chains to accommodate the two functions. In principle, the dual-function RF transmitter may be designed {\em ad hoc} by resorting to suitable waveforms and beamforming strategies, and the resources allocated to the two functions are determined based on the required quality of service. A far less costly strategy relies on the exploitation of existing communication or radar emissions as signals of opportunity to implement the other function without changing the underlying RF transmitter or requiring any additional physical resource.

For example, the signal emitted by TV/FM towers, cellular base stations, and Wi-Fi access points can be used to implement a passive radar, which has the merit of being low-cost, difficult to jam, easy to deploy, and undetectable~\cite{Kuschel-2019}. Interestingly, the {\em opportunistic} radar architectures proposed in~\cite{Nuria-2017,GLV-opportunistic,GLV-opportunistic-adaptive,GLV-2021-TWC} are a practical and effective form of passive radar, as the radar receive chain is placed in close proximity with the existing millimeter-wave (mmWave) communication transmitter and knows its radiated waveform and timing. In particular, the authors in~\cite{Nuria-2016} argue that the opportunistic use of mmWave communication signals is the only credible means to support massive automotive sensing.
 
Radar-enabled communications (i.e., the use of existing radar signals to establish communication links) are less explored.  Even though the idea was proposed back in the 1940's~\cite{Stockman-1948},  systematic studies on the matter have emerged only in the past two decades. For example,~\cite{Hounam-2001} has suggested that targets in synthetic aperture radar images can be tagged for unambiguous identification and localization by equipping them with a  radio-frequency (RF) transponder that downconverts, encodes, and retransmits the received probing signal; the proposed scheme assigns to each target a unique Golay code, which modulates consecutive radar pulses. Also,~\cite{Blunt-2010,Blunt-2011,Metcalf-2015} have investigated covert communications embedded in radar reverberation (clutter). Relying on knowledge of the radar signal and making suitable assumptions on the clutter process, a  transponder undertakes an {\em ad hoc} remodulation of a single radar pulse; the transponder consists of a complete RF receive/transmit chain and operates on a bandwidth wider than that of the radar to create the necessary degrees of freedom.

The idea of exploiting stray signals also underlies the so-called \emph{ambient backscatter} communications, that rely on low-cost and low-power modulators~\cite{Liu13ambientbackscatter}. An ambient backscatterer (hereafter referred to as a tag) is a device that uses existing RF wireless signals as a carrier to communicate with a receiver (hereafter referred to as a reader). Unlike other scattering objects present in the environment, a tag has the ability to alter and reflect the incident signal. Accordingly, ambient backscatter communications can be more power-efficient than traditional radio communications. Existing works show that the signals broadcast from communication sources, e.g., TV/FM towers, cellular base stations, and Wi-Fi access points,  can be exploited for short range data transmission in IoT applications (see~\cite{Huynh2018} and references therein). A major challenge is that the presence of the ambient carrier and the variation of its strength over time may be unknown to both tag and reader, whereby the link quality-of-service may not be entirely under their control; also, the direct interference  from the RF source may be much stronger than the backscattered message. To mitigate theses issues, suitable encoding strategies matching the statistical nature of the ambient signal need to be devised~\cite{Tellambura-2016,Qian-2017a,Qian-2017b, Yang-2018a,Nawaz-2021}, possibly coupled with the exploitation of multiple antennas~\cite{Parks-2014}. In cooperative/cognitive systems, the reader may even jointly decode the messages received from both the RF source and the tag to achieve enhanced spectrum- and energy-efficiency~\cite{Yang-2018b,Zhang-2020}. Reconfigurable intelligent surfaces (RISs) have also proven effective~\cite{Bhowal-2021,Shaoe-2022,BackCom-RIS-2022}; indeed, an RIS can directly act as a tag to provide combined space/time modulation or as a helper to boost the signal strength along the source-tag-reader channel and/or mitigate the interference level along the source-reader link.  On a parallel side, \cite{Itay-2015-journal} has proposed to use a radar transmitter as an ambient source; in particular, it considers a radar base station employing a frequency-modulated continuous-wave signal for ranging multiple sensor nodes and activate their built-in backscatter modulator for data transfer; here, it is assumed that each node only receives the direct radar signal (i.e., the possible echoes from the surrounding objects are neglected); also, since the backscatter receiver is collocated with the radar transmitter, it has knowledge of the radar waveform and timing.

The focus of this paper is on ambient backscatter communications using the radar reverberation  as a carrier signal, as shown in Fig.~\ref{fig_1}. A radar transmits here a periodic signal, which may be for example a low duty cycle pulse train or a modulated continuous wave. Any object located inside the scene illuminated by the radar transmitter inevitably produces scattering in all directions~\cite{Nathanson-book,Skolnik-book}: for example, densely populated areas generate an overwhelming ground clutter, i.e., reflections from static (or almost static, if the wind effect is taken into account) objects, such as walls, buildings, vegetation, man-made infrastructures, mountains, and so on. A tag immersed in clutter on a continuous or almost so basis is endowed with a ``natural'' carrier signal that can be modulated for conveying information towards a reader. The radar clutter hitting the tag may or may not include the direct radar signal, depending on whether the line-of-sight radar-tag  link is clear or obstructed, respectively. The advantages of exploiting both the direct radar signal and the indirect echoes are evident, as the tag may harvest a larger energy over the radar period and backscatter a message even when not directly illuminated by the radar. Similarly, the reverberation generated by the radar transmitter hits the reader, whereby a signal-dependent interference is superimposed on the message arriving from the tag. Hence, the radar clutter is here both a friend (as it provides an ambient carrier at the transmit side) and a foe (as it causes interference at the receive side). 

An important point here is that the clutter received by both the tag and the reader---when observed on conveniently short time intervals---reproduces the {\em periodic structure} of the radar signal. In what follows, we exploit this feature and formulate the problem of designing and assessing radar-enabled backscatter communications under two key constraints:
\begin{itemize}
    \item The backscatter modulator of the tag does not use any RF processing chain, but can only vary the phase/amplitude of the impinging signal at a pre-determined rate.
    \item Tag and reader are synchronized and know the radar period and the channel coherence time; however, in sharp contrast to~\cite{Itay-2015-journal}, they have no information on the radar waveform and the corresponding impinging clutter (i.e., no channel state information is available for the radar-tag-reader and the radar-reader links), so that coordination with the radar transmitter is not required.   
\end{itemize}
Under this scenario, the contributions of the present study can be summarized as follows:
\begin{itemize}
\item We elaborate a convenient signal model and illustrate the interplay among the key system parameters, such as the bandwidth and the period of the radar waveform, the symbol rate employed by the modulator, the duty-cycle of the control signal driving the tag switch, the bandwidth and duration of the receive filter, and the sampling rate. In particular, we show that the unknown baseband pulses carrying the data symbols in the backscattered signal present a periodic structure induced by the radar excitation.

\item We propose encoding/decoding strategies which use a discrete set of reflection coefficients at the tag and are resilient against the radar interference hitting the reader. In particular, we present a promising scheme coupling binary orthogonal coding through the columns of a Hadamard matrix with a differential phase shift keying (PSK) modulation. 

\item We show that the proposed setup carries over plainly to the situation where simultaneous transmissions from multiple synchronous tags must be guaranteed with either a sourced or an unsourced multiple access strategy.

\item Finally, we provide some illustrative examples to assess the effectiveness of the proposed signaling schemes and show some achievable tradeoffs among transmission rate, error probability, and number of supported tags.
\end{itemize}

The potential applications of this form of ambient backscatter communications are many. In particular, a multitude of tags deployed in the region covered by a radar (either indoors or outdoors) can exchange data among them or connect to a central infrastructure with no additional electromagnetic emission and a limited coordination. Thanks to the advances in electronic technologies, radars are becoming more and more accessible to the broad consumer market. Since they work without revealing any information on the personal identities, they can also be mounted in offices, houses, and hospitals and integrated with future terrestrial communication and IoT networks for developing new services based on information about human behavior. The increasing interest in radar functions (ranging from target detection, classification and identification to false alarm control, tracking, and high-accuracy localization) is expected to produce an increment of the number of radar RF transmitters, thus multiplying the opportunities for radar-enabled backscatter communications. Practical use cases may include the exploitation of ground-based radars for air/road traffic control, environmental/weather monitoring, collision avoidance, intrusion detection, and radio imaging.

The remainder of the paper is organized as follows. Sec.~\ref{SEC:System-description} contains the system description and the signal model. Sec.~\ref{SEC:encoding/decoding} presents the encoding/decoding schemes. Sec.~\ref{SEC:Multiple-tag} considers the presence of multiple tags. Sec.~\ref{SEC:Numerical analysis} contains the performance analysis. Finally, Sec.~\ref{SEC:Conclusions} gives some concluding remarks, while the Appendix contains some analytical proofs.

\paragraph*{Notation} In the following, $\mathbb Z$ and $\mathbb C$ are the set of integer and complex numbers, respectively. Column vectors and matrices are denoted by lowercase and uppercase boldface letters, respectively. The symbols $\Re\{\,\cdot\,\}$, $(\,\cdot\,)^{*}$,  $(\,\cdot\,)\transp$, and $(\,\cdot\,)\herm$ denotes real part, conjugate, transpose, and conjugate-transpose, respectively. $\bm{1}_{M}$ and $\bm{0}_{M}$, are the $M$-dimensional all-one and all-zero column vectors, respectively. $a_i$ and  $\|\bm{a}\|$ are the $i$-th entry and the Euclidean norm of the vector $\bm{a}$. $\|\bm{A}\|_F$, $\text{Rank}\{\bm{A}\}$, $\text{Tr}\{\bm{A}\}$, and $\bm{A}^{\dag}$ are the Frobenius norm, the rank, the trace, and the pseudoinverse of the matrix $\bm{A}$. $\bm{I}_{M}$ is the $M\times M$ identity matrix. $\chi_k^2$ denotes the chi-squared distribution with $k$ degrees of freedom. $\chi_k^2(\lambda)$ denotes the noncentral chi-squared distribution with $k$ degrees of freedom and noncentrality parameter $\lambda$. $I_k(\,\cdot\,)$ is the modified Bessel function of the first kind and order $k$. Finally, $\i$, $\star$, and $\text{E}[\,\cdot\,]$ denotes the imaginary unit, the convolution operator, and the statistical expectation, respectively.
 
\section{System description and model development}\label{SEC:System-description}
\begin{figure}[t]
 \centerline{\includegraphics[width=0.8\columnwidth]{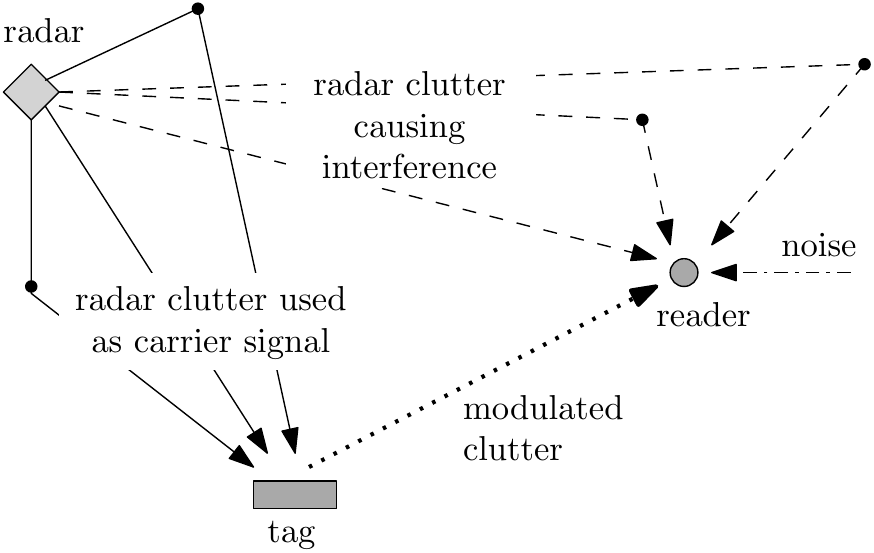}}
 \caption{Graphical illustration of the proposed system: the tag employs the radar clutter as a carrier signal, while the reader aims to decode the message sent by the tag in the presence of the interference (clutter) generated by the radar transmitter.}
 \label{fig_1}
\end{figure}
We consider the ambient backscatter communication system in Fig.~\ref{fig_1}. Here, a radar transmitter illuminates a given region, and passive scatterers produce a reverberation towards both the tag and the reader (including the possible direct signal from the radar). The tag exploits the incident clutter as an ambient carrier to send a message to the reader.  The radar emits the passband signal $\Re\{a(t)\e^{\i 2\pi f_{a}t}\}$, where $f_{a}$ is the carrier frequency and  $a(t)$ is a baseband periodic waveform  of period $T_{a}$ and bandwidth $W_{a}$, so that the radar delay resolution is $1/W_{a}$~\cite{Van-Trees-III}. For a non-scanning (pulsed or continuous-wave) radar, $T_{a}$ is the period of the modulating signal or a multiple thereof; instead, for a scanning radar, it is the scan-time or a multiple thereof. The knowledge of $a(t)$ is not needed in the following developments.

\subsection{Signal emitted by the tag}
Let $\Re\{c(t)\e^{\i 2\pi f_{a}t}\}$ be the radar clutter hitting the tag, where $c(t)$ is its baseband representation; then, we have
\begin{equation}
c(t)=\int_{-\infty}^{\infty}\gamma(t,\tau) a(t-\tau)d\tau
\end{equation}
where $\gamma(t,\tau)$ is the unknown baseband impulse response of the radar-tag channel. We assume that the coherence time of $\gamma(t,\tau)$ spans several, say $L_{a}$, radar periods, so that $c(t)$ is a \emph{locally periodic} signal: this means that $c(t)$  presents $L_{a}$ approximately-equal cycles within any time segment of length $L_{a}T_{a}$, as a consequence of the periodic structure of the radar excitation. For a stationary scenario, $\gamma(t,\tau)=\gamma(\tau)$, and $c(t)$ is itself periodic with period $T_{a}$.

\begin{figure*}[tp]
 \centerline{\includegraphics[width=0.9\textwidth]{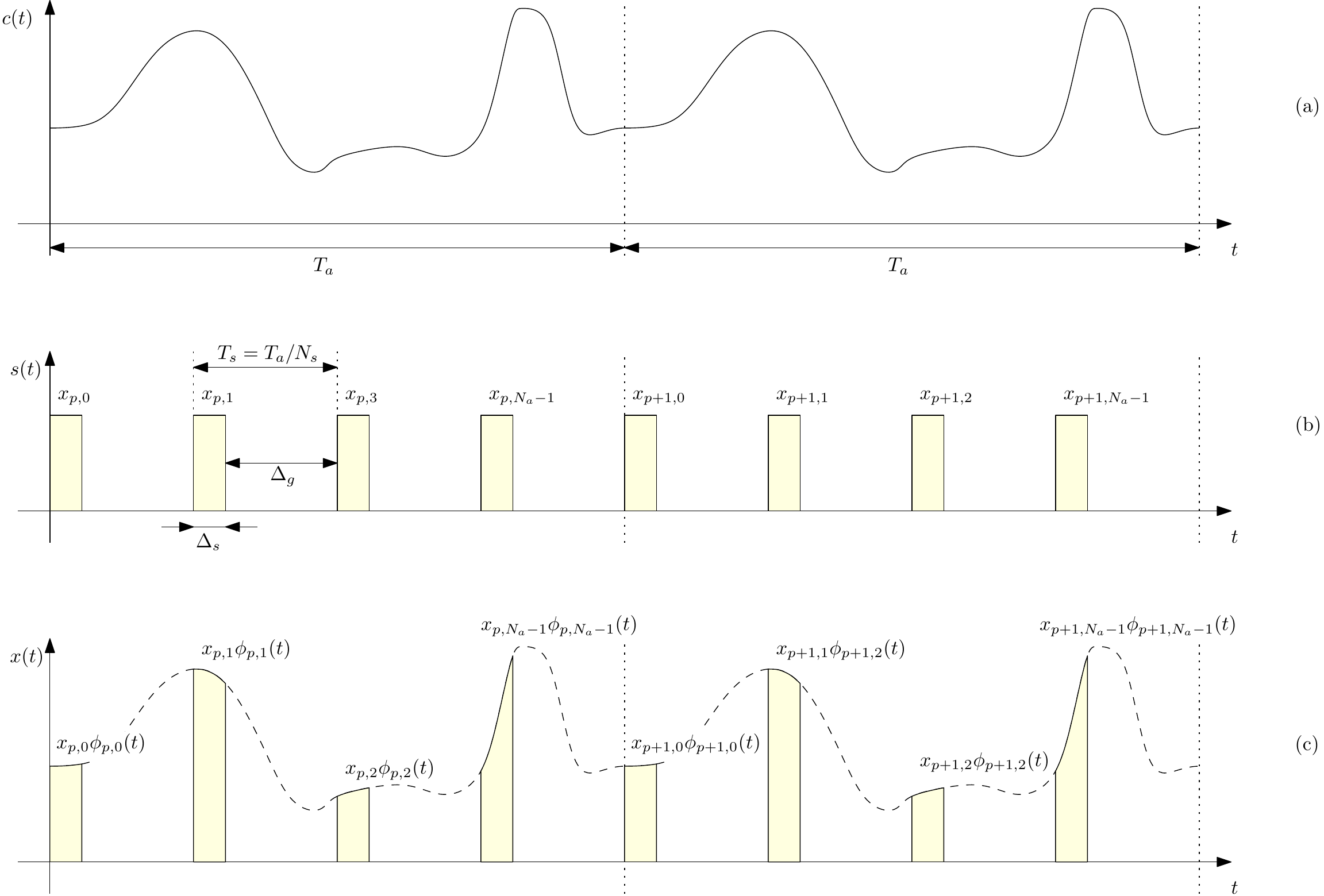}}
 \caption{(a) Radar clutter $c(t)$ employed by the tag as an ambient carrier over two radar periods when $L_{a}\geq 2 $. (b) Message $s(t)$ sent by the tag. (c) Modulated signal $x(t)$ back-scattered by the tag. }
 \label{fig_2}
\end{figure*}

The tag modulates the RF signal $\Re\{c(t)\e^{\i 2\pi f_{a}t}\}$ to send a message $s(t)$~\cite{Huynh2018}; in particular, the backscatter modulator alters this incident waveform by switching the antenna load between two or more states (thus changing its phase and/or amplitude). Assume that the message consists of $N_{s}$ symbols every radar period, so that the symbol interval is $T_{s}=T_a/N_{s}$; also let $x_{p,n}$ be the $n$-th symbol in the $p$-th radar period. Then, the baseband representation of the RF signal backscattered by the tag can be written as\footnote{We neglect the internal thermal noise of the tag, as its circuits consist only of passive components~\cite{Tellambura-2016}.}
\begin{align}
	x(t)&=c(t)\underbrace{\sum_{p\in\mathbb Z} \sum_{n=0}^{N_{s}-1}x_{p,n} \Pi\left(\frac{t-(pN_{s}+n)T_s}{\Delta_{s}}\right)}_{s(t)}\notag \\&=\sum_{p\in\mathbb Z} \sum_{n=0}^{N_{s}-1}x_{p,n} \underbrace{c(t) \Pi\left(\frac{t-(pN_{s}+n)T_s}{\Delta_{s}}\right)}_{\phi_{p,n}(t)}
	 \label{tag-signal}
	\end{align}
where $\Pi(t/\Delta_{s})$ is a rectangular pulse with unit amplitude, support $[0,\Delta_{s}]$, and bandwidth $W_s= 1/\Delta_s$. We assume that $T_s=\Delta_s+\Delta_g$, where $\Delta_g$ is a guard interval between two consecutive transmissions (more on this in Sec.~\ref{SEC_reader_signal}).  Notice that the signal in~\eqref{tag-signal} only accounts for the antenna mode scattering of the tag, which can be varied by acting on the impedance of the antenna load~\cite{BackCom-RIS-2022};\footnote{Since the structural mode scattering does not carry any information, it can be absorbed into the radar interference received by the reader: more on this in Sec~\ref{SEC_reader_signal}.} in particular, $x_{p,n}$ is tied to the complex reflection coefficient induced by the antenna load during the time interval $\big[(pN_{s}+n)T_{s},(pN_{s}+n)T_{s}+\Delta_s\big]$ and belongs to a given discrete alphabet tied to the available hardware (more on this in Sec.~\ref{SEC:encoding/decoding}). Instead, during the silent interval $\big((pN_{s}+n)T_{s}+\Delta_s, (pN_{s}+n+1)T_{s}\big)$ the antenna load is matched to the antenna impedance to avoid signal reflection\footnote{In the case of an imperfect impedance matching, the unmodulated reflected signal produced by the antenna mode scattering can be absorbed into the radar interference received by the reader (as for the structural mode scattering).}~\cite{BackCom-RIS-2022};  batteryless tags can exploit the silent intervals to harvest the energy necessary to run the internal circuitry. A graphical description of $c(t)$, $s(t)$, and $x(t)$ is reported in Fig.~\ref{fig_2}. The following remarks are now in order.

\begin{remark}
The tag performs a temporal gating of the ambient carrier through a switch commuting between the transmit and silent states and loads the symbol $x_{p,n}$ on the pulse $\phi_{p,n}(t)$. The tag has no information on the radar signal and no control on the environmental response. What matters is that, since $c(t)$ is a locally periodic signal, $\phi_{p,n}(t-mT_{a})\simeq\phi_{p+m,n}(t)$ for $m=0,\ldots,L_{a}-1$, whereby two symbols spaced $mN_{s}$ positions apart can be assumed to modulate the same (even though unknown) pulse: see Fig.~\ref{fig_2}(c) for a graphical description.
\end{remark}

\begin{remark}\label{Remark-band}
	If $\Delta_s> 1/W_{a}$, the transmission of $x_{p,n}$ occurs over a time interval larger than the delay resolution of the radar; in this case, the pulse $\phi_{p,n}(t)$ may result from the linear superposition of multiple echoes with resolvable delays that hit the tag over the time interval $\big[(pN_{s}+n)T_{s},(pN_{s}+n)T_{s}+\Delta_s\big]$. If $\Delta_s< 1/W_{a}$, we can instead write
\begin{equation}
	x(t)\simeq\sum_{p\in\mathbb Z} \sum_{n=0}^{N_{s}-1}x_{p,n} \underbrace{c\big((pN_{s}+n)T_s\big) \Pi\left(\frac{t-(pN_{s}+n)T_s}{\Delta_{s}}\right)}_{\phi_{p,n}(t)}
\end{equation}
whereby the pulse $\phi_{p,n}(t)$ approximately maintains an unknown constant amplitude $c\big((pN_{s}+n)T_s\big)$; if $T_s\ll 1/W_{a}$, the value of $c\big((pN_{s}+n)T_s\big)$ may even remain approximately constant over few consecutive transmissions.
\end{remark}

\begin{remark}
The parameters $N_{s}$, $\Delta_s$, and $\Delta_g$ and the alphabet of the tag are under the designer's control. For example, reducing the number $N_{s}$ of symbols per radar period may allow to increase the duration $\Delta_s$ of each modulated pulse and therefore its energy, thus extending the communication range, and/or  
the duration $\Delta_g$ of the guard intervals, thus increasing the amount of energy harvested by the tag. It is understood that batteryless tags can operate only if the harvested energy exceeds a given sensitivity threshold~\cite{Lu-2018,Jia-2022}; accordingly, there is an inherent tradeoff among the number of symbols per radar period and the communication range, which depends upon the power intensity of the incident clutter. Finally, while only a binary alphabet is usually available in low-cost devices~\cite{Huynh2018}, tags equipped with a more advanced hardware could employ more complex modulation formats; in particular, RIS-based tags can implement combined modulation and passive beamforming~\cite{Shaoe-2022,BackCom-RIS-2022}.
\end{remark}

\subsection{Signal received by the reader}\label{SEC_reader_signal}
If the reader knows the radar carrier frequency, its baseband received signal can be written as
\begin{equation}
\tilde y(t) = \beta x(t-\tau) + i(t) + \omega(t) 
\label{rx_signal_0}
\end{equation}
where $\tau\geq 0$ is the tag-reader propagation delay, $\beta\in\mathbb C$ is the unknown attenuation in the tag-reader channel\footnote{To simplify exposition, the tag-reader channel is modeled as a linear-time invariant filter with impulse response $\beta \delta(t-\tau)$, where $\delta(t)$ is the Dirac delta function: for example, this is reasonable when tag and reader are sufficiently close and in line-of-sight. This assumption can however be relaxed.} (including the scattering efficiency of the tag and any carrier phase offset), $i(t)$ is the unknown interference (i.e., the baseband radar clutter hitting the reader, including the structural mode scattering of the tag), and $ \omega(t)$ is the thermal noise, which is modeled as a white complex Gaussian process with power spectral density $\sigma^2_{\omega}$. Even though $c(t)$ and $i(t)$ are produced by the environment in response to the same radar excitation, they are in general different, since tag and reader are in different locations (whereby an echo generated by the same object may arrive with a different delay and/or amplitude) and their antennas have a different orientation and radiation pattern (whereby they may observe echoes originated from different objects). We assume that the coherence time of the radar-reader channel  spans at least $L_{a}$ radar periods, so that $i(t)$ is also locally periodic over any time segment of length $ L_{a}T_{a}$.

The amount of signal power that can be transferred from the tag to the reader depends on the position, orientation, and directivity of their antennas. While position and orientation may in principle be optimized based on some prior cognition on the surrounding environment and the radar location, the beampattern shape is tied to the built-in antenna. Intuitively, while the tag should be fully immersed in clutter, the reader would better be in direct visibility from the tag and, possibly, in a region not reached by a strong radar reverberation. 

\begin{figure*}[t]
 \centerline{\includegraphics[width=0.9\textwidth]{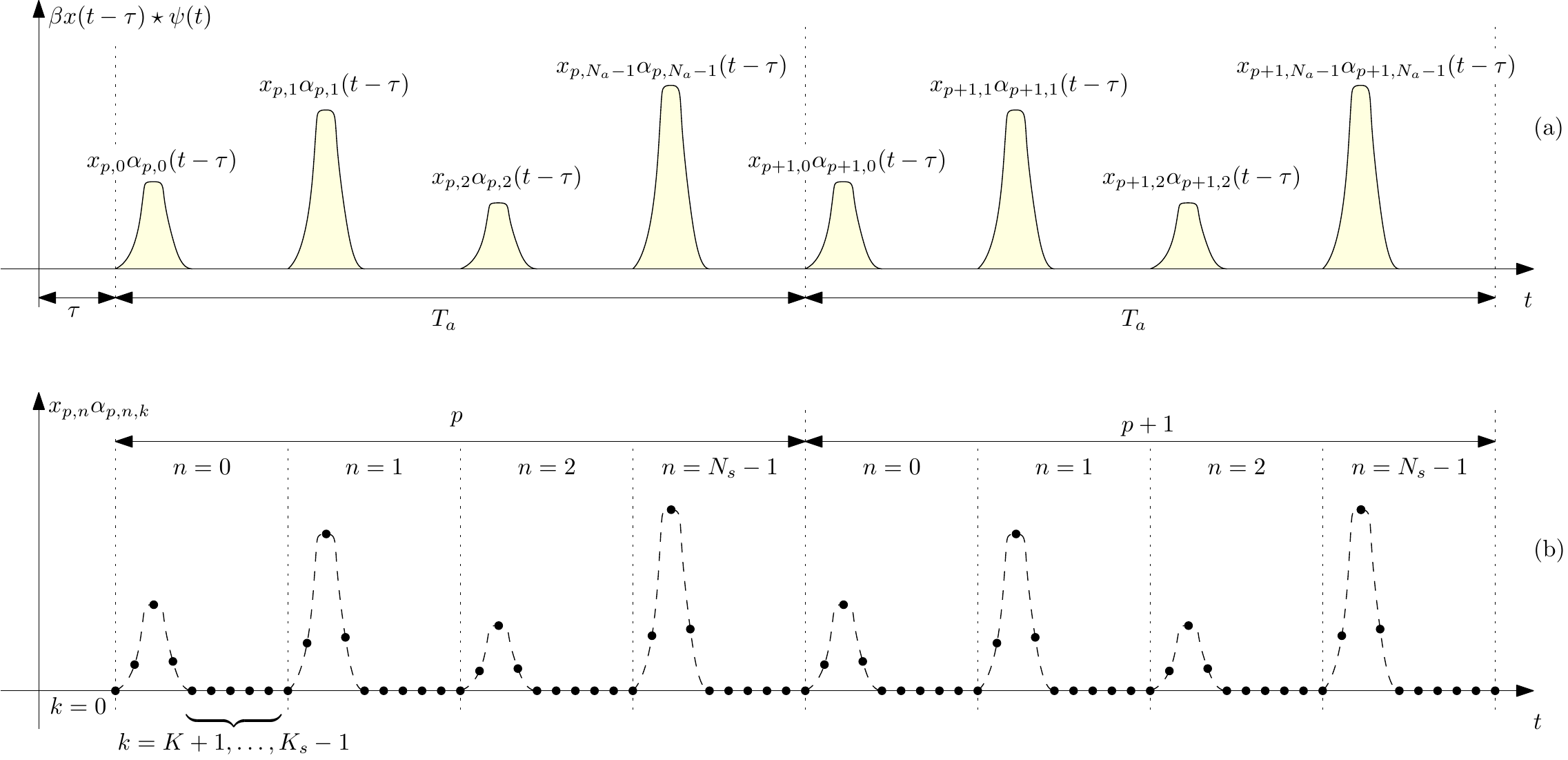}}
 \caption{Example of the waveform $\beta x(t-\tau)\star \psi (t)$ at the output of the receive filter over two radar periods (a) and of the corresponding samples (b) when $L_a\geq 2$, $N_{s}=4$, $K_{s}=9$, and $K=3$.}
 \label{fig_3}
\end{figure*}

The signal in~\eqref{rx_signal_0} is passed through a unit-energy low-pass filter $\psi(t)$, which has bandwidth $W_{\psi}$ and support in $[0,\Delta_\psi]$, with $\Delta_\psi\leq \Delta_g$, whose output is 
\begin{align}
y(t) &= \tilde y(t) \star \psi(t) \notag \\ &=\underbrace{ \sum_{p\in\mathbb Z} \sum_{n=0}^{N_{s}-1}x_{p,n} \underbrace{\beta \phi_{p,n}(t-\tau) \star \psi(t)}_{\alpha_{p,n}(t-\tau)}}_{\beta x(t-\tau)\star \psi (t)} \notag \\ & \quad + i(t)\star \psi (t) + \omega(t) \star \psi (t).
 \label{rx_signal_1}
\end{align}
In the previous equation, $\alpha_{p,n}(t-\tau)$ is the unknown received pulse carrying the symbol $x_{p,n}$, which accounts for the radar clutter $\phi_{p,n}(t-\tau)$ hitting the tag in the time interval $\big[(pN_{s}+n)T_{s},(pN_{s}+n)T_{s}+\Delta_s\big]$, the attenuation $\beta$ of the tag-reader channel, and the receive filter $\psi(t)$ of the reader. The assumption $\Delta_\psi\leq \Delta_g$ implies that $\Delta_s + \Delta_\psi\leq T_s$; accordingly, there is no intersymbol interference in the filtered signal $y(t)$; to better illustrate this point, we provide in Fig.~\ref{fig_3}a a graphical description of the waveform $\beta x(t-\tau)\star \psi (t)$: it is seen that the pulses $\{\alpha_{p,n}(t-\tau)\}$ do not overlap in time.

The signal $y(t)$ is then sampled at rate $K_s/T_s$, where the positive integer $K_s$ is a design parameter. We assume here that tag and reader are synchronized, so that the reader knows when the transmission starts and the channel delay $\tau$; for example, this could be obtained if the
tag is able to receive a reference signal periodically sent by the reader. Accordingly, the sample taken at the epoch $\tau+pT_{a}+nT_{s}+k T_s/K_s$ is
\begin{align}
 y_{p,n,k} &= y(t)\big|_{t=\tau+pT_{a}+nT_{s}+k T_s/K_{s}} \notag\\
 &= x_{p,n} \underbrace{\alpha_{p,n}(t-\tau) \big|_{t=\tau+pT_{a}+nT_{s}+k T_s/K_s}}_{\alpha_{p,n,k}} \notag\\
 & \quad + \underbrace{i(t) \star \psi(t)\big|_{t=\tau+pT_{a}+nT_{s}+k T_s/K_s}}_{i_{p,n,k}}  \notag\\
 & \quad + \underbrace{ \omega(t)\star\psi(t)\big|_{t=\tau+pT_{a}+nT_{s}+kT_s/K_s}}_{ \omega_{p,n,k}}\notag\\
 &= x_{p,n}\alpha_{p,n,k} + i_{p,n,k} + \omega_{p,n,k}\label{rx_signal_2}
\end{align}
for $p\in\mathbb{Z}$, $n=0,\ldots,N_{s}-1$, and $k=0,\ldots,K_s-1$. Fig.~\ref{fig_3}b provides an illustration of the samples of the waveform $\beta x(t-\tau)\star \psi (t)$ in Fig.~\ref{fig_3}a. The following remarks are now given.
\begin{remark} 
To reject the out-of-band noise and preserve the signal of interest, the bandwidth $W_{\psi}$ of the receive filter should equal the bandwidth $W_{x}\simeq W_{a}+W_{s}$ of the  signal $x(t)$ modulated by the tag.\footnote{Notice that $c(t)$ has bandwidth $W_{a}$, as the possible Doppler spread induced by environment is much smaller than $W_{a}$. Hence, the bandwidth of $x(t)=c(t)s(t)$ mainly depends on the tag's symbol rate. If $W_{s}<W_a$, its bandwidth substantially remains in the other of $W_a$, otherwise, increases and is in the other of $W_s$.} Also,  $K_s$ should be at least equal to $\lceil W_{\psi}T_{s}\rceil$ to avoid information loss in the discretization process. Needless to say, lower values of $W_{\psi}$ and/or $K_s$ can be employed to reduce complexity at the price of some loss.
\end{remark}
\begin{remark}\label{Remark-noise-correlation}
The correlation between the noise samples $\omega_{p_{1},n_{1},k_{1}}$ and $\omega_{p_{2},n_{2},k_{2}}$ is
\begin{multline}
	\text{E}\left[\omega_{p_{1},n_{1},k_{1}}\omega^{*}_{p_{2},n_{2},k_{2}}\right]\\=\sigma^2_{\omega}R_{\psi}\Big(\!(p_{1}\!-\!p_{2})T_a\!+\!(n_{1}\!-\!n_{2})T_s\!+\!(k_{1}\!-\!k_{2})T_s/K_s\!\Big)
	\end{multline}
where $R_{\psi}(t)=\psi(t)\star\psi^{*}(-t)$ is the autocorrelation function of $\psi(t)$. Since $R_{\psi}(t)$ has support $[0,2\Delta_{\psi}]$, the above statistical expectation is zero if $p_{1}\neq p_{2}$ or $n_{1}\neq n_{2}$ or $|k_{1}-k_{2}|T_s/K_s\geq2\Delta_{\psi}$; for $p_{1}=p_{2}$, $n_{1}=n_{2}$, $k_{1}\neq k_{2}$, and $|k_{1}-k_{2}|T_s/K_s<2\Delta_{\psi}$, the corresponding samples are uncorrelated only if $R_{\psi}(\ell T_s/K_s)=0$ for $|\ell|=1,\ldots, \lceil2\Delta_{\psi}K_s/T_s\rceil-1$. For example, this is the case when $\psi(t)=\sqrt{K_s/T_s}\Pi(t K_s/T_{s})$. For simplicity, hereafter we assume that the received filter is designed to have uncorrelated noise samples.
\end{remark}
\begin{remark} 
Due to the presence of the guard intervals, we have \begin{equation}
	\alpha_{p,n,k}=0,\quad \text{if } k=0 \text{ or } k=K+1,\ldots,K_s
\end{equation}
where $K=\lceil (\Delta_s + \Delta_\psi) K_s/T_s \rceil-1$. Hence, only the samples $y_{p,n,1},\ldots,y_{p,n,K}$ contain the signal of interest in the time interval $\big[\tau+(pN_{s}+n)T_s,\tau+(pN_{s}+n+1)T_s\big]$, as also shown in Fig.~\ref{fig_3}b: in the following, we only process such data samples and ignore the others.
\end{remark}

\begin{remark}\label{Remark:stationary-clutter}
For any  $n$ and $k$, the sequences $\{\alpha_{p,n,k}\}_{p\in\mathbb{Z}}$ and $\{i_{p,n,k}\}_{p\in\mathbb{Z}}$ contain samples of the waveforms $\sum_{p\in\mathbb{Z}}\alpha_{p,n}(t-\tau)$ and $i(t)$, respectively, which are spaced one radar period apart. The local periodicity of $c(t)$ and $i(t)$ implies that both these sequences are \emph{locally constant}, i.e., that $L_{a}$ consecutive elements are approximately equal.
\end{remark}

\section{Encoding/decoding strategies}\label{SEC:encoding/decoding}
\begin{figure*}[t]
 \centerline{\includegraphics[width=0.9\textwidth]{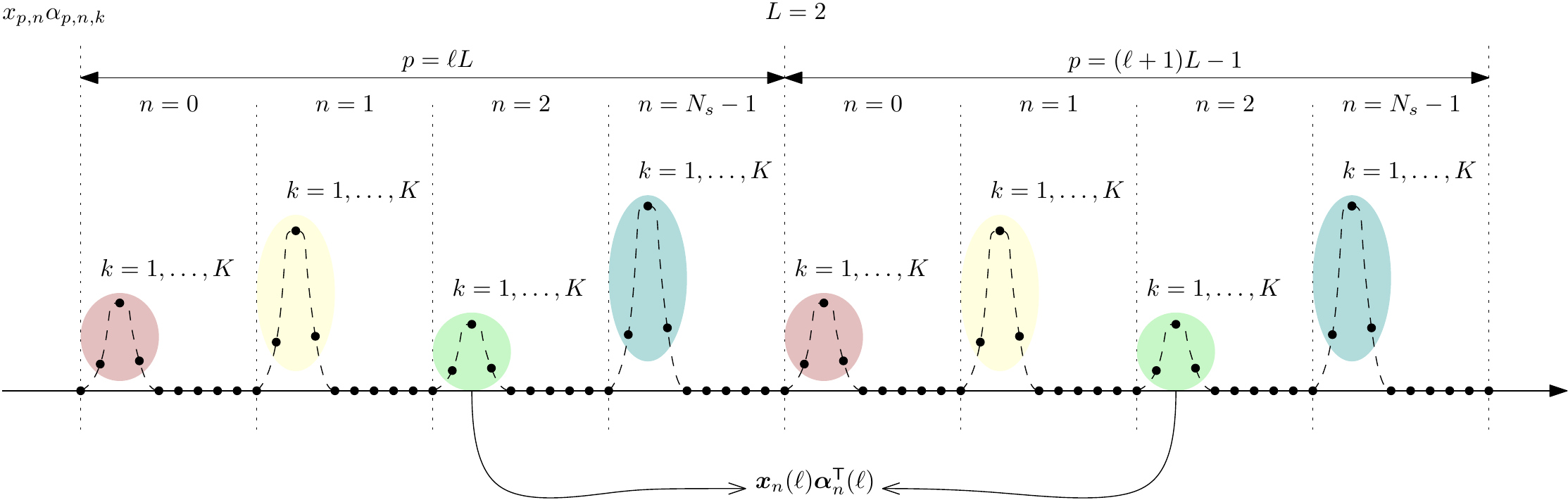}}
 \caption{We consider here a frame spanning $L=2$ radar periods, with $N_{s}=4$, $K_s=9$, and $K=3$, and reports the samples of the waveform $\beta x(t-\tau)\star \psi (t)$ given in Fig.~\ref{fig_3}. Only the non-zero samples corresponding to $k=1,\ldots,K$ are processed; in particular, such samples are parsed in $N_{s}$ groups. Each group corresponds to a different subchannel.}
 \label{fig_4}
\end{figure*}

We propose here encoding/decoding schemes relying only on prior knowledge of the radar period and of the coherence time of the radar-tag-reader and radar-reader channels. We parse the received data samples in~\eqref{rx_signal_2} into $N_{s}$ groups, which define as many time-orthogonal subchannels; the $n$-th subchannel contains the observations $\{y_{p,n,1},\ldots,y_{p,n,K}\}_{p\in\mathbb{Z}}$ taken in the $n$-th symbol interval of each radar period $p\in\mathbb{Z}$, for $n=0,\ldots,N_{s}-1$. This is motivated by the fact that, according to Remark~\ref{Remark:stationary-clutter}, the samples $\alpha_{p,n,1},\ldots,\alpha_{p,n,K}$ of the pulse carrying the symbol $x_{p,n}$ and the corresponding samples $i_{p,n,1},\ldots,i_{p,n,K}$ of the radar interference remain approximately constant in up to $L_{a}$ consecutive uses of the $n$-th subchannel (i.e., $L_{a}$ consecutive radar periods). To take advantage of such memory, we consider disjoint blocks of $L$ consecutive subchannel uses (hereafter referred to as frames), where $L\leq L_{a}$ is a design parameter tied to the affordable system complexity. We assume next that the $\ell$-th frame spans the radar periods indexed by $\ell L,\ldots,(\ell+1) L-1$, with $\ell\in\mathbb{Z}$, and that the reader is aware of the beginning of each frame. In the $\ell$-th frame, the received data samples of the $n$-th subchannels can be organized into the following matrix 
\begin{equation}
\bm Y_{n}(\ell)=	\begin{bmatrix}y_{\ell L,n,1 }&\ldots &  y_{\ell L,n,K} \\
	 \vdots & & \vdots \\
 y_{(\ell+1) L-1,n,1} &\ldots & y_{(\ell+1) L-1,n,K}
 \end{bmatrix}\in\mathbb{C}^{L\times K}.
	\end{equation}
The entries in each row of $\bm Y_{n}(\ell)$ correspond to time epochs spaced $T_s/K_s$ apart within the same radar period (fast-time, in the radar jargon) and contain the same symbol. Instead, the entries in each column of $\bm Y_{n}(\ell)$ correspond to time epochs spaced one radar period apart within the same frame (slow-time, in the radar jargon), so that we can assume here that they share the same value of the carrier signal and radar interference, i.e., 
\begin{subequations}
\begin{align}
	\alpha_{\ell L,n,k} &=\cdots=\alpha_{(\ell+1) L-1,n,k}\\
i_{\ell L,n,k}&=\cdots=i_{(\ell+1) L-1,n,k}
\end{align}
\end{subequations}
for $k=1,\ldots,K$. Hence, upon defining
\begin{subequations}
	\begin{align}
\bm{\alpha}_{n}(\ell)&=\begin{bmatrix}\alpha_{\ell L,n,1 }&\ldots & \alpha_{\ell L,n,K}	\end{bmatrix}\transp\in\mathbb{C}^{K}\\
\bm{i}_{n}(\ell)&=\begin{bmatrix}i_{\ell L,n,1 }&\ldots & i_{\ell L,n,K}	\end{bmatrix}\transp\in\mathbb{C}^{K}\\
\bm{x}_{n}(\ell)&=\begin{bmatrix}x_{\ell L,n }&\ldots & x_{(\ell+1) L-1,n}	\end{bmatrix}\transp\in\mathcal{X}^{L}
\end{align}
\end{subequations}
we can write
\begin{equation}
\bm Y_{n}(\ell)=\bm{x}_{n}(\ell)\bm{\alpha}_{n}\transp(\ell)	+\bm{1}_{L}\bm{i}_{n}\transp(\ell)+\bm{\Omega}_{n}(\ell) \label{rx_signal}
\end{equation}
where $\bm{\Omega}_{n}(\ell)$ is defined similarly to $\bm Y_{n}(\ell)$ and its entries are independent circularly-symmetric Gaussian random variables with variance $\sigma^2_w$. See Fig.~\ref{fig_4} for a graphical description. 

The quality of the $N_{s}$ subchannels may be  different, as they may be sustained by different radar echoes; hence, the tag may decide to use only those experiencing a strong enough ambient carrier. Two strategies are proposed next to use any given subchannel, which rely upon frame-by-frame and frame-differential encoding, respectively; also, we show that joint coding among multiple subchannels is in principle possible.

\subsection{Frame-by-frame encoding}\label{Strategy I}
Consider a given subchannel and frame;  for convenience, we drop here the subchannel index $n$ and the frame index $\ell$, whereby the received signal in~\eqref{rx_signal} becomes
\begin{equation}
\bm Y=\bm{x}\bm{\alpha}\transp+\bm{1}_{L}\bm{i}\transp+\bm{\Omega}. \label{rx_signal_noindex}
\end{equation}
If the codeword $\bm x$ is drawn from a codebook $\mathcal U\subseteq \mathbb C^L$, then the resulting transmission rate is
\begin{equation}
R= \frac{1}{L}\log_2 |\mathcal U|\quad \text{[bits/subchannel-use]}.
\end{equation}
Notice that the transmission of any codeword $\bm u=(u_1\,\cdots\,u_{L})\transp \in {\mathcal U}$ is accomplished by  sending the symbols $u_1,\ldots,u_{L}$ over $L$ consecutive subchannel uses; accordingly, 
the alphabet (i.e., the set of possible reflection coefficients) employed by the tag  is $\mathcal X = \bigcup_{\bm u \in {\mathcal U}} \bigcup_{j=1}^L \{u_j\}$.

The reader is faced with the problem of estimating the codeword $\bm x\in\mathcal U$ when both $\bm \alpha$ and $\bm i$ are unknown and treated as nuisance parameters. It is shown in Appendix~\ref{ML_appendix_multi} that the maximum likelihood (ML) estimator of $\bm x$ is
\begin{equation}
 \hat{\bm x} =\argmax_{\bm u \in \mathcal U} \; \frac{\bigl\Vert \bm u\herm \bm P\bm Y \bigr\Vert^2}{\Vert \bm P \bm u \Vert^2} \label{ML_dec_rule_0}
\end{equation}
where
\begin{equation}
 \bm P =\bm I_L -\frac{1}{L} \bm 1_L \bm 1_L\transp \label{orth_proj_P}
\end{equation}
is the projector onto the orthogonal complement of the subspace spanned by the radar interference.\footnote{The rule in~\eqref{ML_dec_rule_0} is akin to the one employed by a moving target indicator (MTI) radar that estimates the non-zero Doppler shift $\nu$ of a moving target, whose temporal steering vector is $\bm{x}=(1\;\e^{-\i 2 \pi \nu T_{a}}\;\cdots\; \e^{-\i 2 \pi \nu (L-1) T_{a}})\transp$,  while rejecting stationary clutter at zero Doppler frequency.} When the decoding rule in~\eqref{ML_dec_rule_0} is employed, any fraction of power allocated in the subspace spanned by the interference is wasted, as a consequence of the multiplication of $\bm Y$ by $\bm P$; furthermore, any two codewords $\bm{u},\bm{z}\in\mathcal U$ such that $\bm P\bm{u}\propto \bm P\bm{z}$ cannot be distinguished, as a consequence of the fact that the channel state $\bm \alpha$ is unknown. Accordingly, it is desirable to choose a codebook ${\mathcal U}$ possessing the following properties
\begin{enumerate} [leftmargin=1.5cm,label={(P\arabic*)}]
\item \label{P1} $\bm u\herm \bm 1_L = 0$ for any $\bm{u}\in{\mathcal U}$
\item \label{P2} $\text{Rank}\big\{[\bm P\bm{u}\;\;\bm P\bm{z}]\}=2$ for any $\bm{u},\bm{z}\in{\mathcal U}$ and $\bm{u}\neq\bm{z}$
\end{enumerate}
to make the most of the available energy, while keeping all of the codewords distinguishable in the absence of noise. Hereafter, we assume that both \ref{P1} and~\ref{P2} hold, which implies $L\geq 3$ if\footnote{In Sec.~\ref{Strategy II}, we may also allow $|\mathcal U|=1$; in this case only \ref{P1} need to be satisfied, which implies $L\geq 2$.} $|\mathcal U|\geq 2$; accordingly, the decoding rule in~\eqref{ML_dec_rule_0} reduces to
\begin{equation}
 \hat{\bm x} =\argmax_{\bm u \in \mathcal U} \; \frac{\bigl\Vert \bm u\herm \bm Y \bigr\Vert^2}{\Vert \bm u \Vert^2} \label{ML_dec_rule_1}
\end{equation}
so that the interference-free direction (i.e., the codeword) in ${\mathcal U}$ containing the largest portion of the received energy is selected; the rule in~\eqref{ML_dec_rule_1} further simplifies
to \begin{equation}
 \hat{\bm x} =\argmax_{\bm u \in {\mathcal U}} \; \bigl\Vert \bm u\herm \bm Y \bigr\Vert^2. \label{ML_dec_rule}
\end{equation}
if all the codewords have the same energy.  We discuss next two relevant examples.

\begin{example}[Orthogonal codewords]\label{EX-1}
A practical option is to choose a codebook containing orthogonal codewords with equal energy. Without loss of generality, assume that $\|\bm{u}\|^2=L$ for any $\bm{u}\in\mathcal{U}$. In this case, condition~\ref{P1} implies that $|\mathcal U|\leq L-1$, while the decision statistic in~\eqref{ML_dec_rule} takes the form
\begin{equation}
 \bm u\herm \bm Y = \begin{cases}
 L\bm \alpha\transp + \bm w\transp  , & \text{if $\bm x=\bm u$}\\
 \bm w\transp, & \text{otherwise}
 \end{cases}
\end{equation}
where $\bm w\transp= \bm u\herm \bm \Omega$ is a complex circularly-symmetric Gaussian vector with covariance matrix $L\sigma^{2}_{\omega} \bm{I}_{K}$. It is seen that this encoding/decoding strategy is able to get rid of the radar interference and achieve a coherent integration gain of $L$. Also, upon exploiting the fact that
\begin{equation}
 \frac{2}{L\sigma^2_w} \Vert \bm u_j \herm \bm Y \Vert^2 \sim \begin{cases} \chi_{2K}^2 (2L \Vert \bm \alpha \Vert^2/\sigma^2_w), & \text{if } j = i\\
 \chi_{2K}^2, &\text{otherwise}
 \end{cases}\label{Pe_stat}
\end{equation}
if $\bm x = \bm u_i$ is sent, it can be verified that the error probability  conditioned upon $\bm \alpha$ is
\begin{align}
 P\bigl(e \;|\; \bm \alpha \bigr) & = 1 - \int_0^\infty \left( 1- \e^{-x/2} \sum_{k=0}^{K-1} \frac{(x/2)^k}{k!} \right)^{|\mathcal U|-1} \notag\\
 &\quad \times \frac12 \e^{-(x+2L \Vert \bm \alpha \Vert^2/\sigma^2_w)/2}  \left( \frac{x}{2L \Vert \bm \alpha \Vert^2/\sigma^2_w}\right)^{(K-1)/2} \notag\\
 &\quad \times I_{K-1} \left( \sqrt{2L \Vert \bm \alpha \Vert^2 x/\sigma^2_w}\right) dx .\label{Pe_orth}
\end{align}
For $K=1$, the expression in~\eqref{Pe_orth} simplifies to the well known formula
\begin{equation}
 P\bigl(e \;|\; \bm \alpha\bigr) =\sum_{k=1}^{|\mathcal U|-1} \frac{(-1)^{k+1}}{k+1} \binom{|\mathcal U|-1}{k} \e^{- \frac{k}{k+1} \frac{2L \Vert \bm \alpha \Vert^2}{2\sigma^2_w}}\label{Pe_FSK}
\end{equation}
that is the error probability of orthogonal signaling with non-coherent detection~\cite{Proakis-book}. If $L$ is a power of 2, a possible choice for $ \mathcal U$ are the columns (except $\bm 1_L$) of the Hadamard matrix of order $L$, so that $\mathcal X = \{-1, 1\}$, and only $2$ phase states are needed at the tag.
\end{example}

\begin{example}[$M$-PSK alphabet] If the entries of the codewords in $\mathcal U$ are taken from the $M$-PSK alphabet $\{\e^{\i 2\pi m/M}\}_{m=0}^{M-1}$, then we have $\mathcal{X}=\{\e^{\i 2\pi m/M}\}_{m=0}^{M-1}$, and only $M$ phase states are needed at the tag.  Tables~\ref{tab_3} and~\ref{tab_4} show the number of codewords, the transmission rate (in bits/subchannel-use) and the worst-case cosine similarity among the codewords 
\begin{equation}
		\max_{\substack{ \bm{u},\bm{z}\in\mathcal{U}\\\bm{u}\neq\bm{z}}}\frac{\big|\bm{u}\herm \bm{z}\big|}{\|\bm{u}\|\|\bm{z}\|} \label{cos_sim}
\end{equation}
for the largest codebook satisfying~\ref{P1} and~\ref{P2}. Only non-trivial designs with at least two codewords are shown. For $M>2$, the above quantities have been obtained via an exhaustive search. For $M=2$, instead, it can be shown that~\ref{P1} and~\ref{P2} implies that $L$ must be even and limits the number of codewords to $\frac 12 \binom{L}{L/2}$;  in this case, the transmission rate is $R=\frac{1}{L}\log_2 \bigl(\frac 12 \binom{L}{L/2}\bigr)$, the worst-case cosine similarity is\footnote{This result follows from the facts that, in~\eqref{cos_sim}, $\Vert \bm u \Vert = \Vert \bm z \Vert =\sqrt{L}$, $|\bm u\herm \bm z| = \bigl| \sum_{i=1}^L u_i z_i\bigr| = |2 \delta - L|\leq L- 4$, where $\delta$ is the number of positions in which $u_i=z_i$, and equality holds if $\delta = 2$ or $\delta =L-2$.} $| 1- 4/L|$, and both increase with $L$. For comparison, Tables~\ref{tab_3} and~\ref{tab_4} also analyze the relevant case where the additional constraint that the codewords be mutually orthogonal is enforced; in this case, the transmission rate obtained with the largest possible codebook inevitably converges to zero with an increasing $L$.  
\end{example}

\begin{table}[t]
 \centering 
 \caption{Number of codewords, transmission rate (in bits/subchannel-use) and worst-case cosine similarity among the codewords for the largest codebook satisfying~\ref{P1} and~\ref{P2} when a $2$-PSK alphabet is used \label{tab_3}}
 \begin{tabular}{ccccccc}
 \toprule
 \multirow{2}[1]{*}{$L$} & \multicolumn{2}{c}{No. codewords} & \multicolumn{2}{c}{Rate} & \multicolumn{2}{c}{Similarity} \\
 \cmidrule(lr){2-3} \cmidrule(lr){4-5} \cmidrule(lr){6-7}& non-orth. & orth. & non-orth. & orth. & non-orth. & orth.\\
 \midrule
 4 & 3 & 3 & 0.3962 & 0.3962 & 0 & 0 \\
 6 & 10 & $-$ & 0.5537 & $-$ & 0.3333 & $-$\\
 8 & 35 & 7 & 0.6412 & 0.3509 & 0.5000 & 0 \\
 10 & 126 & $-$ & 0.6977 & $-$ & 0.6000 & $-$ \\
 12 & 462 & 11 & 0.7907 & 0.2883 & 0.6667 & 0 \\
 14 & 1716 & $-$ & 0.7675 & $-$ & 0.7143 & $-$\\
 16 & 6435 & 15 & 0.7729 & 0.2442 & 0.7500 & 0\\
 18 & 24310 & $-$ & 0.8094 & $-$ & 0.7778 & $-$\\
 \bottomrule
 \end{tabular}
\end{table}

\begin{table}[t]
 \centering 
 \caption{Number of codewords, transmission rate (in bits/subchannel-use) and worst-case cosine similarity among the codewords for the largest codebook satisfying~\ref{P1} and~\ref{P2} when an $M$-PSK alphabet is used \label{tab_4}}
 \begin{tabular}{cccccccc}
 \toprule
 \multirow{2}[1]{*}{$M$} & \multirow{2}[1]{*}{$L$} & \multicolumn{2}{c}{No. codewords} & \multicolumn{2}{c}{Rate} & \multicolumn{2}{c}{Similarity} \\
 \cmidrule(lr){3-4} \cmidrule(lr){5-6} \cmidrule(lr){7-8}& & non-orth. & orth. & non-orth. & orth. & non-orth. & orth.\\
 \midrule
 \multirow{2}{*}{3} & 3 & 2 & 2 & 0.3333 & 0.3333 & 0.6667 & 0 \\
 & 6 & 30 & 5 & 0.8178 & 0.3870 & 0.8333 & 0 \\
 \midrule
 \multirow{3}{*}{4} & 4 & 9 & 3 & 0.7925 & 0.3962 & 0.7500 & 0\\
 & 6 & 100 & 5 & 1.1073 & 0.3870 & 0.8333 & 0\\
 & 8 & 1225 & 7 & 1.2823 & 0.3509 & 0.8750 & 0\\
 \midrule
 5 & 5 & 24 & 4 & 0.9170 & 0.4 & 0.8000 & 0 \\
 \midrule
 \multirow{6}{*}{6} & 3 & 2 & 2 & 0.3333 & 0.3333 & 0.6667 & 0\\
 & 4 & 15 & 3 & 0.9767 & 0.3962 & 0.8660 & 0\\
 & 5 & 60 & $-$ & 1.1814 & $-$ & 0.8718 & $-$\\
 & 6 & 340 & 5 & 1.4016 & 0.3870 & 0.8819 & 0\\
 & 7 & 1680 & 6 & 1.5306 & 0.3693 & 0.8921 & 0 \\
 & 8 & 9135 & 7 & 1.6446 & 0.3509 & 0.9014 & 0 \\
 \bottomrule
 \end{tabular}
\end{table}

\subsection{Frame-differential encoding}\label{Strategy II}
A major limitation of the signaling scheme in Sec.~\ref{Strategy I} is the shortage of low-correlated codewords satisfying conditions~\ref{P1} and~\ref{P2}; in particular, the number of orthogonal messages is inherently limited by the chosen frame length $L$. When $L\leq L_{a} /2$, we can overcome this drawback by resorting to a more elaborated signaling scheme which includes a differential encoding over two consecutive frames. 

To proceed, we consider a given subchannel;  for convenience, we drop the subchannel index $n$, whereby the received signal in~\eqref{rx_signal} becomes
\begin{equation}
\bm Y(\ell)=\bm{x}(\ell)\bm{\alpha}\transp(\ell)+\bm{1}_{L}\bm{i}\transp(\ell)+\bm{\Omega}(\ell). \label{rx_signal_ell}
\end{equation}
The tag sends the message $\bm x(\ell) = b(\ell) \bm u(\ell)$, where $\bm u(\ell)\in\mathcal U$, and $\{b(\ell)\}_{\ell \in \mathbb Z}$ is a sequence of differentially encoded $M$-PSK symbols, with $M\geq 2$. In this case, the alphabet used by the tag is $\mathcal X =  \bigcup_{m=0}^{M-1} \bigcup_{\bm u \in \mathcal U}\bigcup_{j=1}^L \{\e^{\i 2\pi m/M} u_j\}$, while the transmission rate is 
\begin{equation}
R= \bigl(\log_2 |\mathcal U|+\log_2 M\bigr)/L\quad  \text{[bits/subchannel-use]}.
\end{equation}
Let $\theta(\ell)$ denote the incremental phase shift in the differential encoding, i.e., $b(\ell) = \e^{\i \theta(\ell)} b(\ell-1)$, and consider the observable $\bm Y(\ell)$ and $\bm Y(\ell-1)$ collected in two subsequent frames. Then, since $\bm \alpha(\ell) = \bm \alpha(\ell-1)$ and $\bm i(\ell) = \bm i(\ell-1)$, the ML estimate of $\bigl( \theta(\ell) , \bm u(\ell)\bigr)$ based on $\bm Y(\ell)$ and $\bm Y(\ell-1)$ takes the following form
\begin{multline}
 \bigl(\hat \theta(\ell) , \hat{\bm u}(\ell) \bigr) = \argmax_{\substack{ \theta \in \{2\pi m/M\}_{m=0}^{M-1}\\ \bm u\in \mathcal U}} \max_{\bm z \in \mathcal U} \left\Vert \begin{bmatrix} \e^{\i \theta} \bm u \\ \bm z \end{bmatrix}\herm \begin{bmatrix} \bm Y (\ell) \\ \bm Y(\ell-1) \end{bmatrix} \right\Vert^2\\=\argmax_{\substack{ \theta \in \{2\pi m/M\}_{m=0}^{M-1}\\ \bm u\in \mathcal U}} \max_{\bm z \in \mathcal U} \left\Vert \e^{-\i \theta} \bm u\herm \bm Y(\ell) + \bm z\herm \bm Y(\ell-1) \right\Vert^2.\label{joint_search}
\end{multline}

Notice that a search over a set of cardinality $M|\mathcal U|^2$ is required in~\eqref{joint_search}. To reduce the decoding complexity, we can replace the nuisance parameter $\bm u(\ell-1)$ by its estimate, say $\bm u'(\ell-1)$, obtained in the previous frame; in this case, the decoding rule is
\begin{multline}
 \{\hat \theta'(\ell) , \hat{\bm u}'(\ell) \} =\\\argmax_{\substack{ \theta \in \{2\pi m/M\}_{m=0}^{M-1}\\ \bm u \in \mathcal U}} \left\Vert \e^{-\i \theta} \bm u\herm \bm Y(\ell) + \bigl(\hat{\bm u}'(\ell-1)\bigr)\herm \bm Y(\ell-1) \right\Vert^2\label{disjoint_search}
\end{multline} which requires a search over a set of cardinality $M|\mathcal U|$. An even simpler decoding rule is obtained by relaxing the grid $\{2\pi m/M\}_{m=0}^{M-1}$ to the continuous interval $[0,2\pi]$ and, then, projecting the relaxed solution onto the original search set; in this case, we have
\begin{subequations}\label{rule_combined_3}
\begin{align}
\hat{\bm u}_{\text{sub}}(\ell) &= \argmax_{\bm u \in \mathcal U} \; \bigl\Vert \bm u\herm \bm Y(\ell) \bigr\Vert^2 \label{ML_dec_v}\\
\hat \theta_{\text{sub}}(\ell) & = \!\!\!\argmin_{\theta \in \{\frac{2\pi m}{M}\}_{m=0}^{M-1}} \!\bigl| \theta \!-\! \angle\bigl( \hat{\bm u}_{\text{sub}}\herm (\ell) \bm Y(\ell) \bm Y\herm (\ell\!-\!1) \hat{\bm u}_{\text{sub}} (\ell\!-\!1) \bigr) \!\bigr|\label{ML_dec_Delta}
\end{align}
\end{subequations}
so that the reader first recovers $\bm u(\ell)$ from~\eqref{ML_dec_v} and then accomplishes differential decoding with~\eqref{ML_dec_Delta} by considering the statistic $\hat{\bm u}_{\text{sub}}\herm (\ell) \bm Y(\ell) \bm Y\herm (\ell-1) \hat{\bm u}_{\text{sub}} (\ell-1)$. 

\begin{example}[Combined orthogonal/differential encoding]
A simple scheme is obtained when the set $\mathcal U$ contains mutually-orthogonal and equal energy codewords, as in Example~\ref{EX-1}. When a binary differential encoding is adopted, the possible messages are $\{\bm u, -\bm u\}_{\bm u \in\mathcal{U}}$. Also, if $M$ is a power of 2 and the entries of the codewords in $\mathcal U$ are also taken from an $M$-PSK alphabet, then only $M$ phase states are needed at the tag.
\end{example}

\begin{example}[Combined repetition/differential encoding]
Another practical option is to choose $\mathcal U = \{\bm u\}$, so that the tag sends $\bm{x}(\ell)=b(\ell) \bm{u}$, where $\bm u$ is any vector orthogonal to $\bm{1}_{L}$. In this case, the ML decoding rule in~\eqref{joint_search} becomes
\begin{align}
 \hat \theta(\ell) & = \argmin_{\theta \in \{\frac{2\pi m}{M}\}_{m=0}^{M-1}} \left| \theta - \angle \bigl(\bm u\herm \bm Y (\ell) \bm Y\herm(\ell-1) \bm u \bigr)\right|.
\end{align}
Remarkably, if $M$ and $L$ are even and $\bm u$ is a vector with half entries equal to $1$ and the other half equal to $-1$, only $M$ phase states are needed at the tag.  Also, notice that the error probability is available in closed-form for $K=1$~\cite{Proakis-book}: it is the error probability of the $M$-ary differential phase shift keying modulation; in particular, we have
\begin{multline}
 P\bigl( e \,|\, \alpha(\ell)\bigr)  = \frac{1}{\pi} \int_0^{\pi-\pi/M} \e^{-\frac{\sin^2(\pi/M)}{1+\cos(\pi/M)\cos \theta)} \frac{L |\alpha(\ell)|^2}{\sigma^2_w}} d\theta \\
  \begin{cases} = \frac12 \e^{-L |\alpha(\ell)|^2/\sigma^2_w}, & \!\!\!\!\text{if } M=2\\
 \approx 2 Q\!\left(\!\! \sqrt{\!\frac{L |\alpha(\ell)|^2}{\sigma^2_w} \sin^2(\! \frac{\pi}{M}\!)}\right)\!, & \!\!\!\!\text{if } M>2 \text{ and } L |\alpha(\ell)|^2 \!\gg \!\sigma^2_w
 \end{cases}
\end{multline}
where $Q(x)=\int_x^\infty \frac{1}{\sqrt{2\pi}} \e^{-t^2/2} dt$.
\end{example}

\subsection{Encoding across subchannels}
It is worth noticing that encoding across $B\geq 2$ subchannels can also be  performed. Let $n_1,\ldots,n_B$ denote the indexes of such subchannels; we only discuss next two simple examples, deferring to future studies a more in depth analysis.

\begin{example}[Repetition encoding]
The same message is repeated here over the $B$ subchannels, so that $\bar{\bm x}(\ell)=\bm x_{n_1}(\ell)=\cdots=\bm x_{n_B}(\ell)$. In this case, we can arrange the corresponding received samples in each frame $\ell$ as follows
\begin{align}
\bar{\bm Y}(\ell)&=\begin{bmatrix}\bm Y_{n_1}(\ell)&\cdots&\bm Y_{n_{B}}(\ell)\end{bmatrix}\notag\\
&=\bar{\bm x}(\ell)\bar{\bm{\alpha}}\transp(\ell)	+\bm{1}_{L}\bar{\bm{\imath}}\transp(\ell)+\bar{\bm{\Omega}}(\ell) \in \mathbb{C}^{L\times BK}
\label{RX_signal_sub_repetition}
\end{align}
where $\bar{\bm{\alpha}}(\ell)=[\bm{\alpha}_{n_1}\transp(\ell)\cdots\bm{\alpha}_{n_B}\transp(\ell)]\transp$, $\bar{\bm{\imath}}(\ell)=[\bm{i}_{n_1}\transp(\ell)\cdots\bm{i}_{n_B}\transp(\ell)]\transp$, and $\bar{\bm{\Omega}}(\ell)=[\bm{\Omega}_{n_1}(\ell)\cdots\bm{\Omega}_{n_B}(\ell)]$. The signaling strategies illustrated in the Secs.~\ref{Strategy I} and~\ref{Strategy II} can be directly applied to the input/output model in~\eqref{RX_signal_sub_repetition}. Pros and cons are evident: indeed, an energy and, possibly, a diversity gain can be obtained at the price of a rate loss. Accordingly, this strategy  may be helpful when the considered subchannels are sustained by weak or fluctuating echoes.
\end{example}

\begin{example}[Subchannels with equal clutter]
Assume to have $B$ adjacent subchannels experiencing the same carrier signal and radar interference, so that $\bar{\bm{\alpha}}(\ell)=\bm{\alpha}_{n_1}(\ell)=\cdots=\bm{\alpha}_{n_B}(\ell)$ and
$\bar{\bm{\imath}}(\ell)=\bm{i}_{n_1}(\ell)=\cdots=\bm{i}_{n_B}(\ell)$. This occurs when $BT_s$ is much smaller than the radar delay resolution. In this case, we can arrange the received samples as 
\begin{align}
\bar{\bm Y}(\ell)&=\begin{bmatrix}\bm Y_{n_1}\transp(\ell)&\cdots&\bm Y_{n_{B}}\transp(\ell)\end{bmatrix}\transp\notag\\
&=\bar{\bm x}(\ell)\bar{\bm{\alpha}}\transp(\ell)	+\bm{1}_{L}\bar{\bm{\imath}}\transp(\ell)+\bar{\bm{\Omega}}(\ell) \in \mathbb{C}^{BL\times K}
\label{RX_signal_wideband_tag}
\end{align}
where $\bar{\bm{x}}(\ell)=[\bm{x}_{n_1}\transp(\ell)\cdots\bm{x}_{n_B}\transp(\ell)]\transp$ and $\bar{\bm{\Omega}}(\ell)=[\bm{\Omega}_{n_1}\transp(\ell)\cdots\bm{\Omega}_{n_B}\transp(\ell)]\transp$. Again, the signaling strategies illustrated in Secs.~\ref{Strategy I} and~\ref{Strategy II} can be directly applied to the input/output model in~\eqref{RX_signal_wideband_tag}. The main advantage here is that communication can take place even when $L=1$ by making the bandwidth of the modulated signal much larger than that of the radar excitation.
\end{example}

\section{Multiple tags}\label{SEC:Multiple-tag}
The above signaling schemes can be generalized to serve multiple synchronous tags on the same subchannel. For brevity, we only consider here frame-by-frame encoding.

\subsection{Sourced multiple access}\label{SEC:sourced}
Consider first a sourced multiple access with $Q$ active tags; in this case, the reader is interested in both the received messages and the identities of the tags that generated them. Since all tags are synchronized, their backscattered signals arrive \emph{time-aligned} to the reader, and the model in~\eqref{rx_signal_noindex} can be modified as follows
\begin{align}
\bm Y&=\sum_{q=1}^{Q}\bm{x}^{(q)}\left(\bm{\alpha}^{(q)}\right)\transp	+\bm{1}_{L}\bm{i}\transp+\bm{\Omega}\notag \\&=\bm{X}^{(1:Q)}\left(\bm{A}^{(1:Q)}\right)\transp+\bm{1}_{L}\bm{i}+\bm{\Omega} \label{rx_signal_multi}
\end{align}
where $\bm{x}^{(q)}\in \mathcal{U}^{(q)}$ is the codeword sent by the tag $q$, $\mathcal{U}^{(q)}$ is the codebook employed by the tag $q$, $\bm{\alpha}^{(q)}$ contains the samples of the pulse carrying the symbols in $\bm{x}^{(q)}$, $\bm{X}^{(1:Q)}=\bigl[\bm{x}^{(1)} \cdots \bm{x}^{(Q)}\bigr]$, and 
$\bm{A}^{(1:Q)}=\bigl[\bm{\alpha}^{(1)} \cdots \bm{\alpha}^{(Q)}\bigr]$. As shown in Appendix~\ref{ML_appendix_multi}, the ML decoding rule is now
\begin{equation}
 \hat{\bm X}^{(1:Q)}=\argmax_{\bm U \in \mathcal{U}_{s}^{(1:Q)}} \left\Vert \bm P \bm U (\bm P \bm U)^\dagger \bm P \bm Y \right\Vert^2_F \label{ML_dec_rule_multi}
\end{equation}
where $\bm P$ is given in~\eqref{orth_proj_P}, $\bm P\bm{U}(\bm P \bm U)^\dagger$ is the orthogonal projector on the column space of the matrix $\bm P\bm{U}$, and
\begin{equation}
\mathcal{U}_{s}^{(1:Q)}=\Big\{[\bm{u}^{(1)}\;\cdots\;\bm{u}^{(Q)}]:\; \bm{u}^{(q)}\in \mathcal{U}^{(q)}\;\forall\;q\Big\}. \label{set_UQ}
\end{equation}

When the decoding rule in~\eqref{ML_dec_rule_multi} is employed, any fraction of power allocated in the subspace spanned by the interference is wasted; also, the matrices $\bm{U},\bm{Z}\in\mathcal{U}_{s}^{(1:Q)}$ cannot be distinguished if  $\bm P\bm{U}$ and $\bm P\bm{Z}$ present the same column span, while $\bm{U}$ will always be preferred to $\bm{Z}$ if the column span of $\bm P\bm{Z}$ is strictly contained in that of $\bm P\bm{U}$. Accordingly, it is desirable that the adopted codebooks $\mathcal{U}^{(1)},\ldots,\mathcal{U}^{(Q)}$ possess the following properties
\begin{enumerate} [leftmargin=1cm,label={(P\arabic*s)}]
\item \label{P1s} $\bm U\herm \bm 1_L = \bm{0}_{Q}$ for any $\bm{U}\in\mathcal{U}_{s}^{(1:Q)}$
\item \label{P2s} $\text{Rank}\big\{\bm P\bm{U}\big\}=Q$ and $\text{Rank}\big\{[\bm P\bm{U}\;\bm P\bm{Z}]\}\geq Q+1$ for any $\bm{U},\bm{Z}\in \mathcal{U}_{s}^{(1:Q)}$ and $\bm{U}\neq\bm{Z}$
\end{enumerate}
which generalize those given in Section~\ref{Strategy I}; these
properties ensure that the identities of the tags and their messages are identifiable in the absence of noise and imply $L\geq Q+2$.\footnote{A necessary condition to satisfy \ref{P1s} and \ref{P2s} is that different users employ different codebooks. Indeed, since no channel state information is available, user identification is possible only by assigning them a unique codebook.} 
In this case, the decoding rule in~\eqref{ML_dec_rule_multi} simplifies to
\begin{equation}
 \hat{\bm X}^{(1:Q)}=\argmax_{\bm U \in \mathcal{U}_{s}^{(1:Q)}} \; 
 \bigl\Vert \bm{U}\bm{U}^{\dag}\bm Y_n(\ell)\bigr\Vert^2_F. \label{ML_dec_rule_0_multi}
\end{equation}

The implementation of the  ML rule in~\eqref{ML_dec_rule_0_multi} requires a joint search among $Q$ tags. Interestingly, as summarized in Algorithm~\ref{alg:omp}, we may use here the standard orthogonal matching pursuit (OMP)~\cite{Tropp-2007} to obtain a suboptimal solution to~\eqref{ML_dec_rule_0_multi} which only entails a sequence of $Q$ one-dimensional searches.  At each step of Algorithm~\ref{alg:omp}, the codeword with the largest contribution to the residual signal $\bm Z$, as measured by the norm of the inner product with $\bm Z$, is added to the set of detected codewords. Once a codeword belonging to tag $m_j$ is detected, then the codebook $\mathcal U^{(m_j)}$ of this tag is excluded from the search space in subsequent iterations (see step~\ref{OMP:set-update}). The current set of detected codewords form the columns of a matrix $\hat {\bm X}$ and the residual is computed by subtracting from $\bm Y$ the projection of $\bm Y$ onto $\hat {\bm X}$'s column space. The complexity of Algorithm~\ref{alg:omp} is dominated by the pseudoinverse in step~\ref{OMP:pinv} and is $\mathcal O(QL^3)$.
\begin{algorithm}[t]
 \caption{Orthogonal Matching Pursuit (sourced)}
 \begin{algorithmic}[1]
 \STATE $\mathcal M=\{1,\ldots,Q\}$, $\bm{Z} = \bm Y$
 \FOR{$j = 1,\dots,Q$} 
 \STATE $m_j = \argmax\limits_{q\in\mathcal M} \max\limits_{\bm u\in \mathcal U^{(q)}}\|\bm u^H \bm Z\|^2$
 \STATE $\hat{\bm x}^{(m_j)} = \argmax\limits_{\bm u\in \mathcal U^{(m_j)}}\|\bm u^H \bm Z\|^2$ 
 \STATE \label{OMP:set-update} $\mathcal M = \mathcal M \setminus\{m_j\}$
 \STATE $\hat{\bm X} =[\hat{\bm x}^{(m_1)}\,\cdots\,\hat{\bm x}^{(m_j)}]$
 \STATE \label{OMP:pinv} $\bm Z = \bigl(\bm I_L - \hat{\bm X}\hat{\bm X}^\dag\bigr) \bm Y$
 \ENDFOR
 \RETURN $\hat{\bm x}^{(1)},\ldots,\hat{\bm x}^{(Q)}$
 \end{algorithmic}
 \label{alg:omp}
\end{algorithm}

\subsection{Unsourced multiple access}
Consider now an unsourced multiple access where $Q_{\max}$ active tags employ the same codebook $\mathcal{U}$. The reader is interested only in the received messages, while the identities of the tags that generated them is irrelevant~\cite{liu2022unsourced}. The number $Q$ of  distinct messages may be lower than the number of active tags, as more tags may transmit the same codeword. Let $\mathcal{H}_{Q}$ be the hypothesis that $Q$ distinct messages are selected, for $Q=1,\ldots,Q_{\max}$. Under $\mathcal{H}_{Q}$, the received signal can be still expressed as in~\eqref{rx_signal_multi} with $\bm{X}^{(1:Q)}\in\mathcal{U}_{u}^{(1:Q)}$, where
\begin{multline}
\mathcal{U}_{u}^{(1:Q)}=\big\{[\bm{u}^{(1)}\;\cdots\;\bm{u}^{(Q)}]: \; \bm{u}^{(q)}\in\mathcal{U}\;\forall\; q  \text{ and } \\ \bm{u}^{(i)}\neq\bm{u}^{(j)}\;\forall\;i\neq j\big\}.
\end{multline}
By the same arguments illustrated before, it is desirable that $\mathcal{U}$ satisfies \ref{P1} given in Sec.~\ref{Strategy I}; also, in the absence of noise, the identifiability of the messages under $\mathcal{H}_{Q}$ is possible if 
\begin{enumerate}[leftmargin=1cm, label={(P\arabic*u)}]
\setcounter{enumi}{1}
\item \label{P2u} $\text{Rank}\big\{\bm{U}\big\}=Q$ and $\text{Rank}\big\{[\bm{U}\,\bm{Z}]\}\geq Q+1$ for any $\bm{U},\bm{Z}\in \mathcal{U}_{u}^{(1:Q)}$ and $\bm{U}\neq\bm{Z}$.
\end{enumerate}
Under the above assumptions, the resulting average transmission rate is\footnote{Assume for example that $Q_{\max}=2$ and that each tag randomly selects a codeword from $|\mathcal{U}|$. If the tags select the same codeword, which occurs with probability $1/|\mathcal{U}|$, the received message conveys $\log_2|\mathcal{U}|$ bits in $L$ subchannel uses; if instead the tags select distinct codewords, which occurs with probability $(|\mathcal{U}|-1)/|\mathcal{U}|$, the two received  messages convey $\log_2|\mathcal{U}|+\log_2(|\mathcal{U}|-1)$ bits in $L$ subchannel uses. Accordingly, we have  \begin{equation}\label{eq:unsourced_rate_2}R=\frac{1}{2L|\mathcal{U}|}\log_2|\mathcal{U}|+\frac{|\mathcal{U}|-1}{2L|\mathcal{U}|}\left(\log_2|\mathcal{U}|+\log_2(|\mathcal{U}|-1)\right).\end{equation}}
\begin{equation}
	R= \frac{1}{LQ_{\max}}\text{E}\left[\sum_{q=0}^{Q-1} \log_2(|\mathcal{U}|-q)\right]\; \text{[bits/subchannel-use/tag]}
\end{equation}
where the expectation is over the random variable $Q\in\{1,\ldots,Q_{\max}\}$. Also, upon resorting to a generalized information criterion (GIC), an estimate $\hat{Q}$ of the number of messages is~\cite{stoica-2004-model,GLV-2021-TWC}
\begin{equation}\label{ML_dec_rule_0_unsourced}
\hat{Q}=\argmax_{Q\in\{1,\ldots,Q_{\max}\}} \left\{\max_{{\bm U} \in \mathcal{U}_{u}^{(1:Q)}} \;
\bigl\Vert \bm{U}\bm{U}^{\dag}\bm Y_n(\ell)\bigr\Vert^2_F-\gamma Q\right\}
\end{equation}
where $\gamma$ is a convenient penalty factor.
For a given $\hat{Q}$, the ML estimate of the messages is 
\begin{equation}\label{ML_dec_rule_0_unsourced_b}
\hat{\bm{X}}^{(1:\hat{Q})}=\argmax_{{\bm U} \in \mathcal{U}_{u}^{(1:\hat{Q})}} \; 
\bigl\Vert \bm{U}\bm{U}^{\dag}\bm Y_n(\ell)\bigr\Vert^2_F.
\end{equation}
Alternatively, we may resort to an OMP-based procedure which extracts the superimposed back-scattered messages one-by-one~\cite{GLV-2021-TWC}, as summarized in Algorithm~\ref{alg:omp_us}.

\begin{algorithm}[t]
 \caption{Orthogonal Matching Pursuit (unsourced)}
 \begin{algorithmic}[1]
 \STATE Given number of tags $Q_{\max}$, shared codebook $\mathcal U$, threshold $\eta$
 \STATE $\hat Q=0$, $\bm{Z} = \bm Y$
 \FOR{$j = 1,\dots,Q_{\max}$} 
 \STATE $\hat Q=\hat Q+1$
 \STATE $\hat{\bm x}^{(j)} = \argmax\limits_{\bm u\in \mathcal U}\|\bm u^H \bm Z\|^2$ 
 \STATE $\hat{\bm X} =[\hat{\bm x}^{(1)}\,\cdots\,\hat{\bm x}^{(j)}]$
 \STATE $\bm Z = \bigl(\bm I_L - \hat{\bm X}\hat{\bm X}^\dag\bigr) \bm Y$
 \IF{$\|\bm Z\|_F < \eta$}
 \STATE \textbf{break}    
 \ENDIF
 \ENDFOR
 \RETURN $\hat{\bm x}^{(1)},\ldots,\hat{\bm x}^{(\hat Q)}$ 
 \end{algorithmic}
 \label{alg:omp_us}
\end{algorithm}

\section{Performance analysis}\label{SEC:Numerical analysis}
In this section we provide some examples to assess the performance of the proposed signaling schemes and illustrate some relevant system tradeoffs. We assume that the clutter samples remain constant over a frame and then independently change from frame to frame, thus resulting in a block fading channel model. Following~\cite{Shnidman-1999}, the squared clutter samples have a noncentral chi-square density with two degrees of freedom; more specifically, we assume in~\eqref{rx_signal} that  $\bm{\alpha}_{n}(\ell) = \bm{\alpha}_{s,n}(\ell) + \bm{\alpha}_{d,n}(\ell)$, where 
$\bm{\alpha}_{s,n}(\ell)$ and $\bm{\alpha}_{d,n}(\ell)$ represent the specular (deterministic) and diffuse (random) components, respectively. The entries of $\bm{\alpha}_{s,n}(\ell)$ have the same magnitude $\sigma_{s,\alpha}>0$, while $\bm{\alpha}_{d,n}(\ell)$ is a complex circularly-symmetric Gaussian random vector with covariance matrix $\bm C\in\mathbb{C}^{K\times K}$; the entries of $\bm C$ are modeled as $[\bm C]_{ij}= \rho^{|i-j|}\sigma^{2}_{d,\alpha}$ for some $\rho\in[0,1]$ and $\sigma^{2}_{d,\alpha}>0$. Notice that
$\kappa_{\alpha}=\sigma^{2}_{s,\alpha}/ \sigma^{2}_{d,\alpha}$ is the power ratio between the specular and diffuse components, $(\sigma^{2}_{s,\alpha} + \sigma^{2}_{d,\alpha})/\sigma_w^2$ is the received signal-to-noise (SNR), and $\rho$ rules the covariance among the entries of $\bm{\alpha}_{n}(\ell)$; in particular, sampling the received signal at the Nyquist rate produces low- or highly-correlated entries if $\Delta_s\gg 1/W_{a}$ or $\Delta_s\ll 1/W_{a}$, respectively (see also Remark~\ref{Remark-band}). Unless otherwise stated, we assume $\kappa_{\alpha}=1/9$, $\rho=0$, $L=8$, and $K=2$. Finally, the codewords are assumed equally probable, and the system performance is assessed in terms of $P(e)$, i.e., the probability that a message is erroneously decoded by the reader, which is computed by  averaging over $10^{6}$ channel realizations. 

\subsection{Frame-by-frame encoding}\label{SEC-Permormance-I}
We consider here the signaling strategy discussed in Sec.~\ref{Strategy I}. The codebook $\mathcal{U}$  adopted in~\eqref{rx_signal_noindex} satisfies both~\ref{P1} and~\ref{P2}, with the entries of all codewords taken from a $2$-PSK alphabet. The feasible designs for $L\leq 18$ are listed in Table~\ref{tab_3}; for example, $\mathcal{U}$ can contains up to $35$ codewords when $L=8$, with at most $7$ mutually-orthogonal codewords corresponding to the columns of the $8\times 8$ Hadamard matrix, except the all-one vector.

\begin{figure}[!t]
 \centerline{\includegraphics[width=\columnwidth]{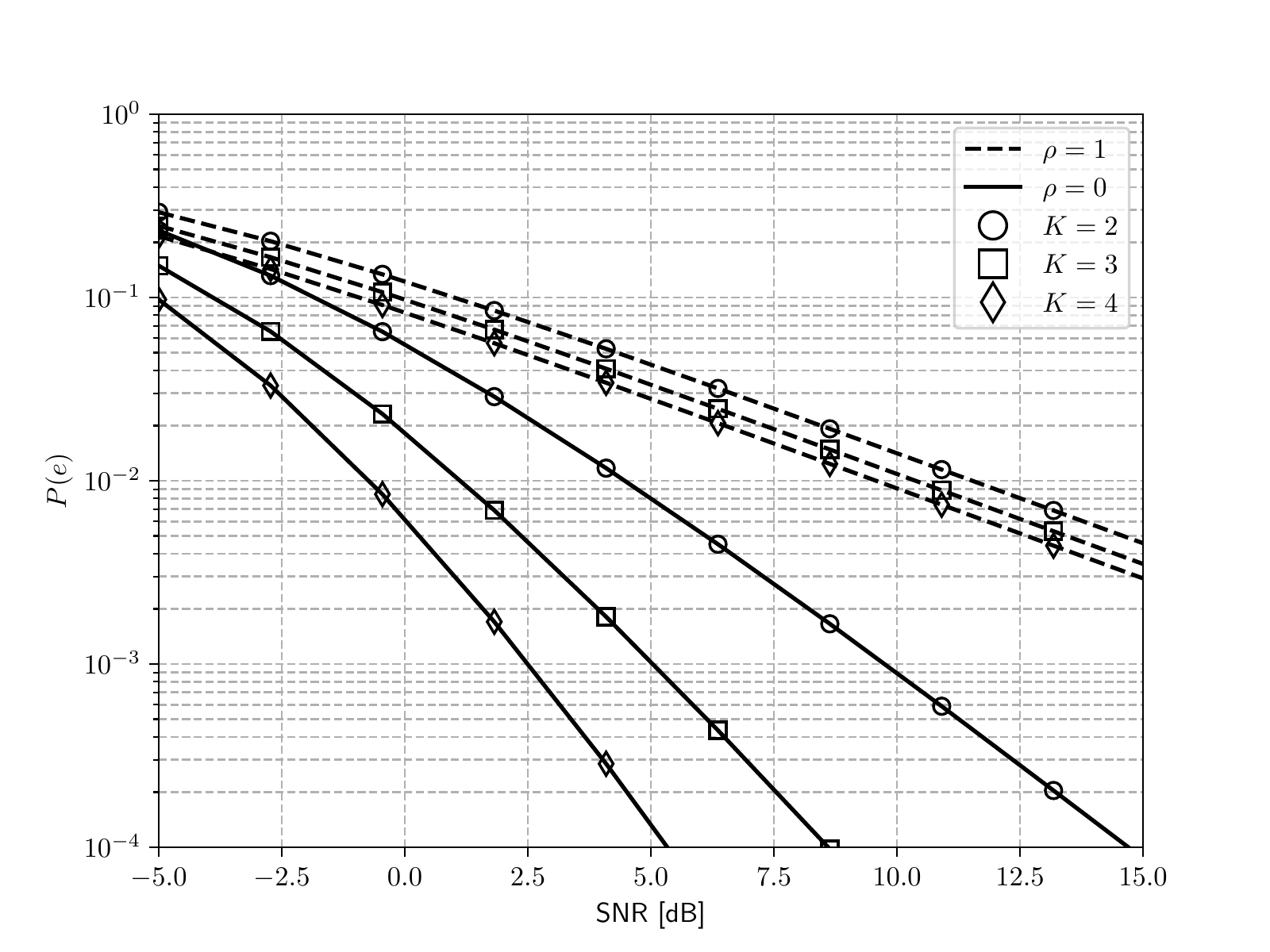}}
 \vspace{-0.2cm}
 \caption{$P(e)$ versus SNR for $K=2,3,4$ and $\rho=0,1$, when $L=8$, $\kappa_{\alpha}=1/9$, the frame-by-frame encoding strategy in Sec.~\ref{Strategy I} is employed, and the codebook contains $4$ mutually-orthogonal codewords (therefore, the transmission rate is $0.25$ bits/subchannel-use). \label{fig:hadamard}}
 
 \centerline{\includegraphics[width=\columnwidth]{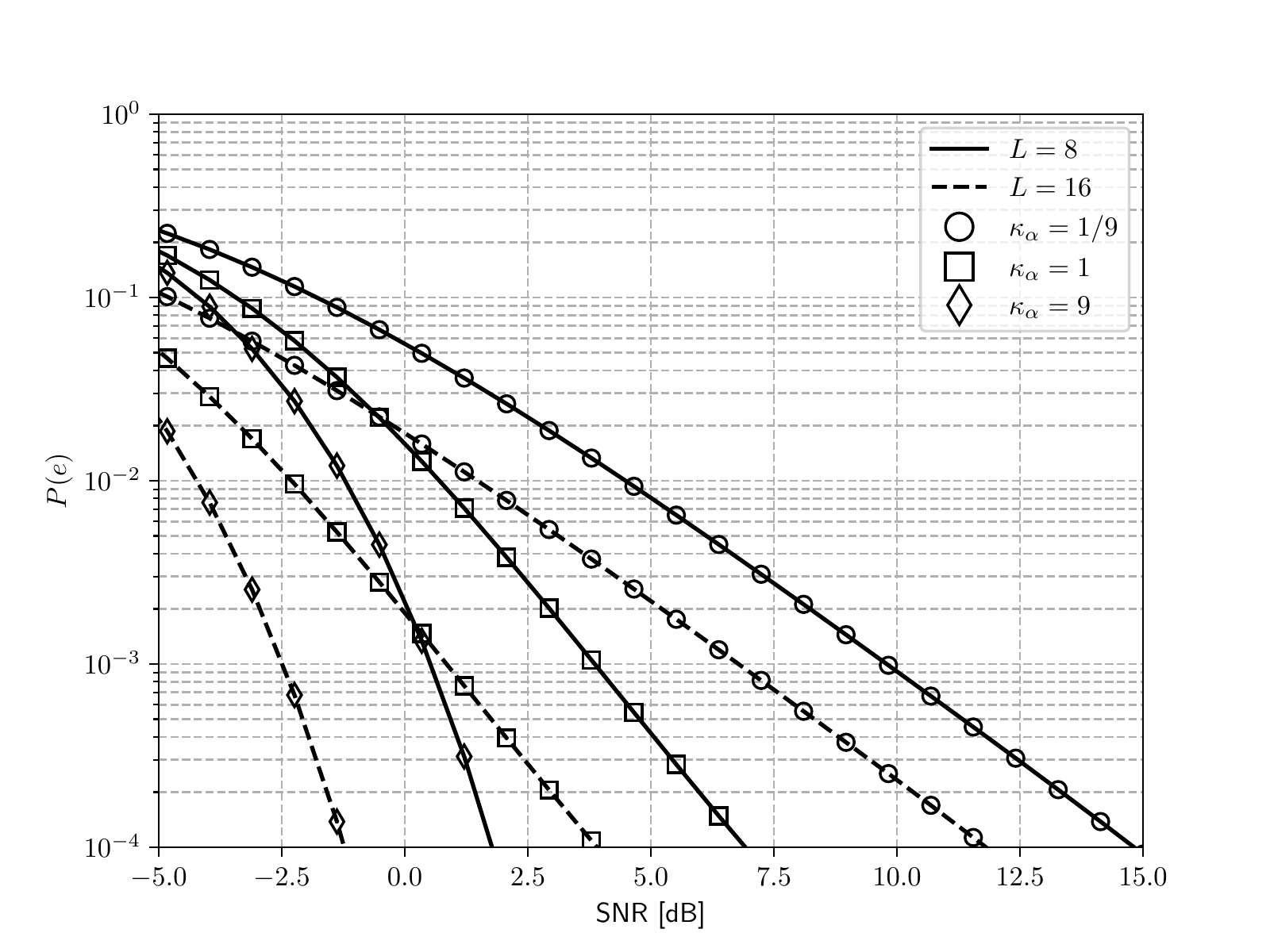}}
  \vspace{-0.2cm}
 \caption{$P(e)$ versus SNR for $\kappa_{\alpha}=1/9,1,9$ and $L=8,16$, when $K=2$, $\rho=0$, the frame-by-frame encoding strategy in Sec.~\ref{Strategy I} is employed, and the adopted codebook contains $4$ mutually-orthogonal codewords (therefore, the transmission rate is $0.25$ and $0.125$ bits/subchannel-use for $L=8,16$, respectively). \label{fig:v0}}
\end{figure}

\begin{figure}[!t]	
 \centerline{\includegraphics[width=\columnwidth]{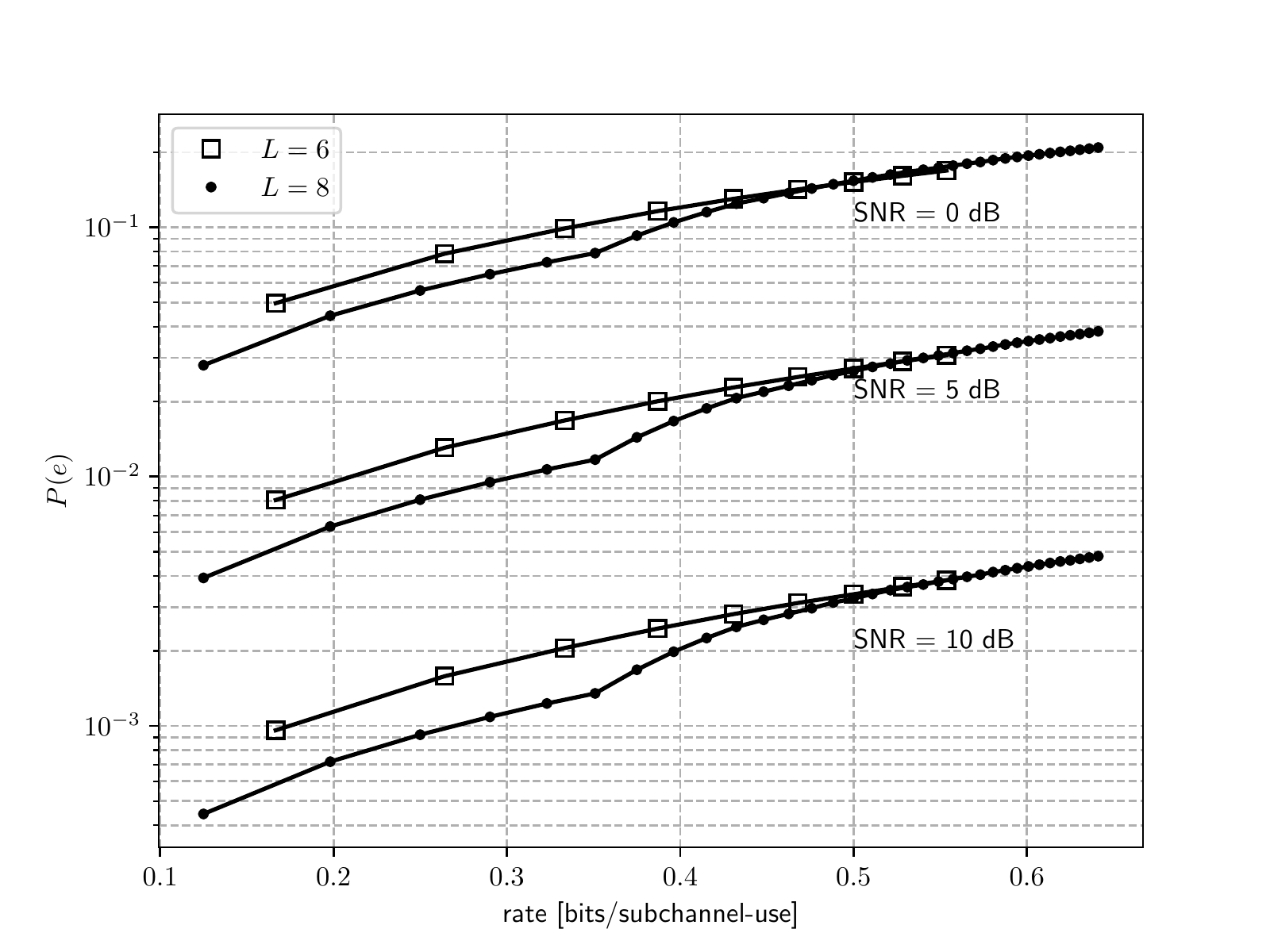}}
  \vspace{-0.2cm}
 \caption{$P(e)$ versus the transmission rate for $\text{SNR}=0,5,10$~dB and $L=6,8$, when $K=2$, $\rho=0$, $\kappa_{\alpha}=1/9$, and the frame-by-frame encoding strategy in Sec.~\ref{Strategy I} is employed.}
 \label{fig:cardU}
\end{figure}

We first assume that $\mathcal{U}$  contains $4$ mutually-orthogonal codewords and study the impact of $K$, $\rho$, $\kappa_{\alpha}$, and $L$. Fig.~\ref{fig:hadamard} shows $P(e)$  versus SNR for $K=2,3,4$ and $\rho=0,1$; it is seen that increasing $K$ (i.e., the duration $\Delta_s$ of each transmitted pulse for a fixed radar bandwidth and sampling rate) provides an SNR gain, as a longer segment of the radar clutter hitting the tag is exploited to convey each symbol; for $\rho=0$, a diversity gain $K$ is also obtained, as we have $K$ uncorrelated clutter samples. Fig.~\ref{fig:v0} shows $P(e)$  versus SNR for $\kappa_{\alpha}=1/9,1,9$ and $L=8,16$; $P(e)$  decreases as $\kappa_{\alpha}$ increases, as a stronger specular component is present; also, there is an integration gain of $3$~dB when $L$ is doubled at the price of a rate loss, in keeping with Example~\ref{EX-1}.

Finally, Fig.~\ref{fig:cardU} shows $P(e)$  versus the transmission rate, for $\text{SNR}=0,5,10$~dB and $L=6,8$. The codebook size is varied from $2$ to $10$ for $L=6$ and from $2$ to $35$ for $L=8$, according to Table~\ref{tab_3}; for an increasing size, the codebook is augmented by first including the mutually-orthogonal codewords (when they exist) and then the non-orthogonal ones. It is seen that $P(e)$  gracefully degrades as the transmission rate increases; also, more favorable tradeoffs are obtained with a larger $L$.

\subsection{Frame-differential encoding}
\begin{figure}[!t]
 \centerline{\includegraphics[width=\columnwidth]{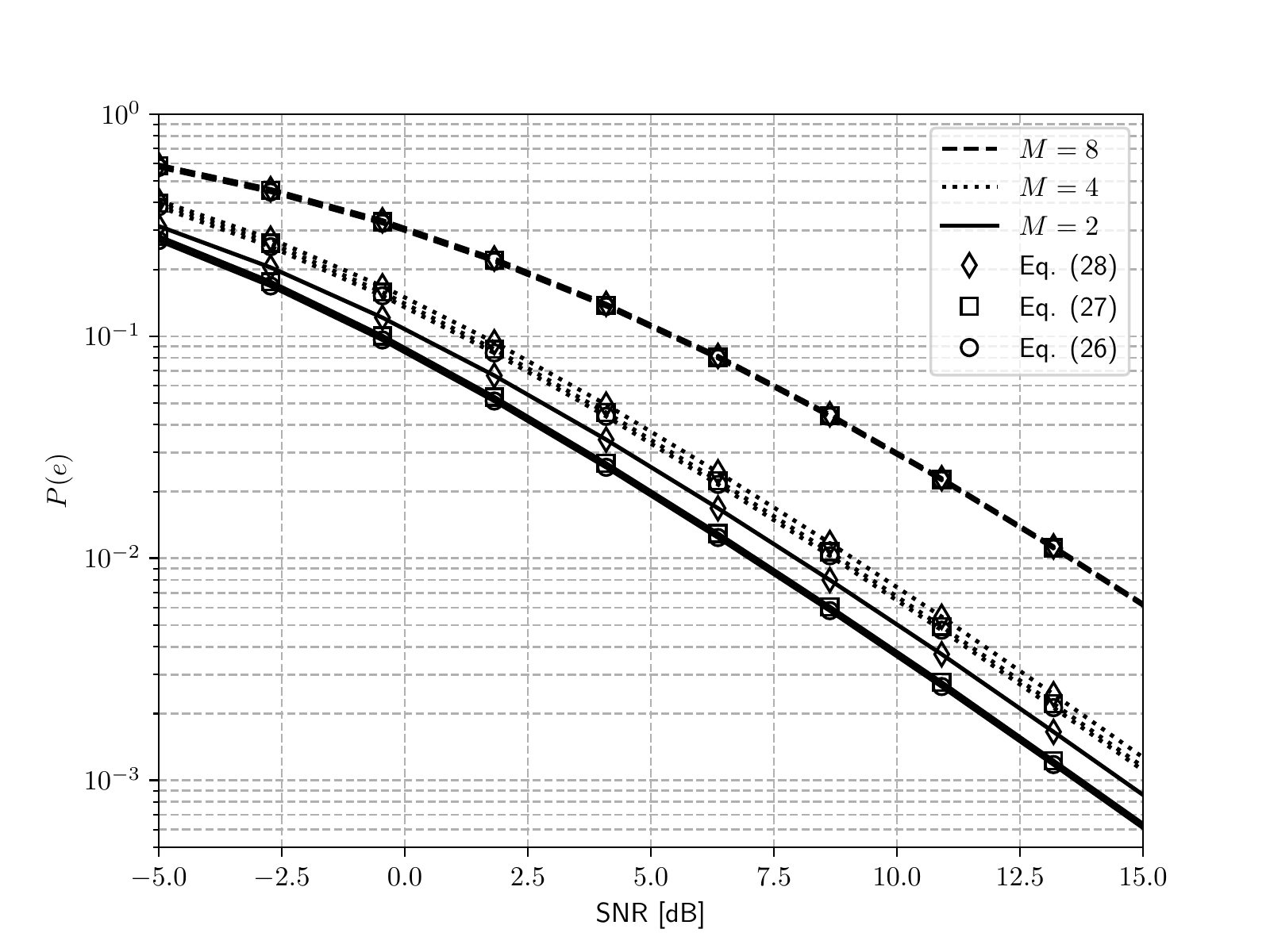}}
 \vspace{-0.2cm}
 \caption{$P(e)$ versus SNR for $M=2,4,8$, when $L=8$, $K=2$, $\rho=0$, $\kappa_{\alpha}=1/9$, the frame-differential encoding strategy in Sec.~\ref{Strategy II} is employed and $\mathcal U$ contains $4$ mutually-orthogonal codewords (therefore, the transmission rate is $0.375,0.5, 0.625$ bits/subchannel-use for $M=2,8$, respectively). \label{fig:differential}} 

\centerline{\includegraphics[width=\columnwidth]{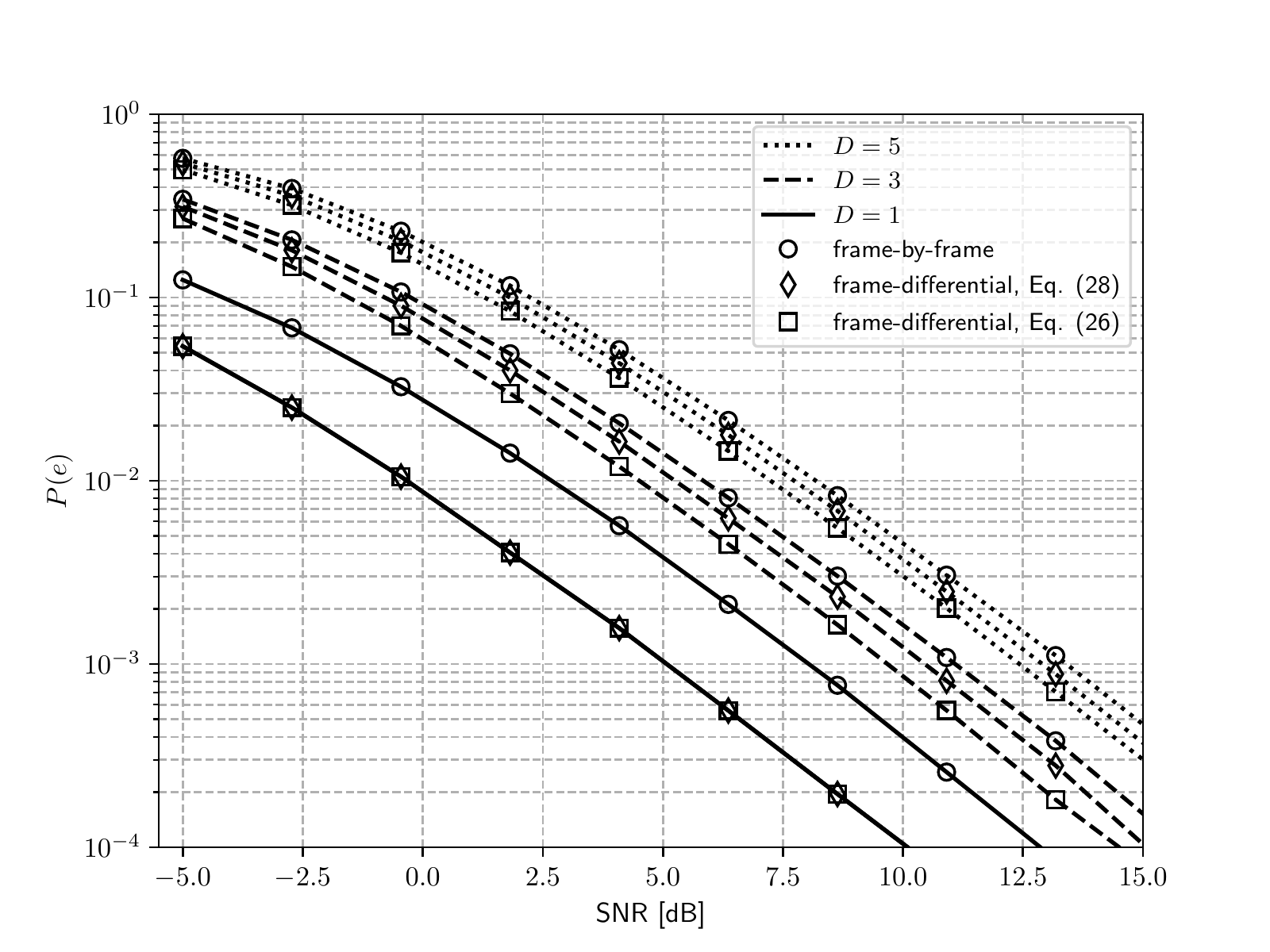}}
 \vspace{-0.2cm}
 \caption{$P(e)$ versus SNR when $L=8$, $K=2$, $\rho=0$, and $\kappa_{\alpha}=1/9$. The transmission rate is $D/L$ bits per subchannel-use, for $D=1,3,5$. For frame-differential encoding, $M=2$ and $|\mathcal{U}|=2^{D-1}$; for frame-by-frame encoding, $|\mathcal{U}|=2^D$. 
 \label{fig:differential-a}}
\end{figure}

We consider here the signaling strategy discussed in Sec.~\ref{Strategy II}; the set $\mathcal{U}$ adopted in~\eqref{rx_signal_ell} is constructed as in Sec.~\ref{SEC-Permormance-I}. In Fig.~\ref{fig:differential}, we first compare the decoding rules in~\eqref{joint_search}, \eqref{disjoint_search}, and~\eqref{rule_combined_3}, when $M=2,4,8$ and $\mathcal{U}$ contains $4$ mutually-orthogonal codewords. It is seen that the first two rules provide a similar error probability; also, the rule in~\eqref{rule_combined_3} is only slightly inferior with respect to the former two and, therefore, can be preferred in practice to reduce complexity. 

\begin{figure}[!t]
\centerline{\includegraphics[width=\columnwidth]{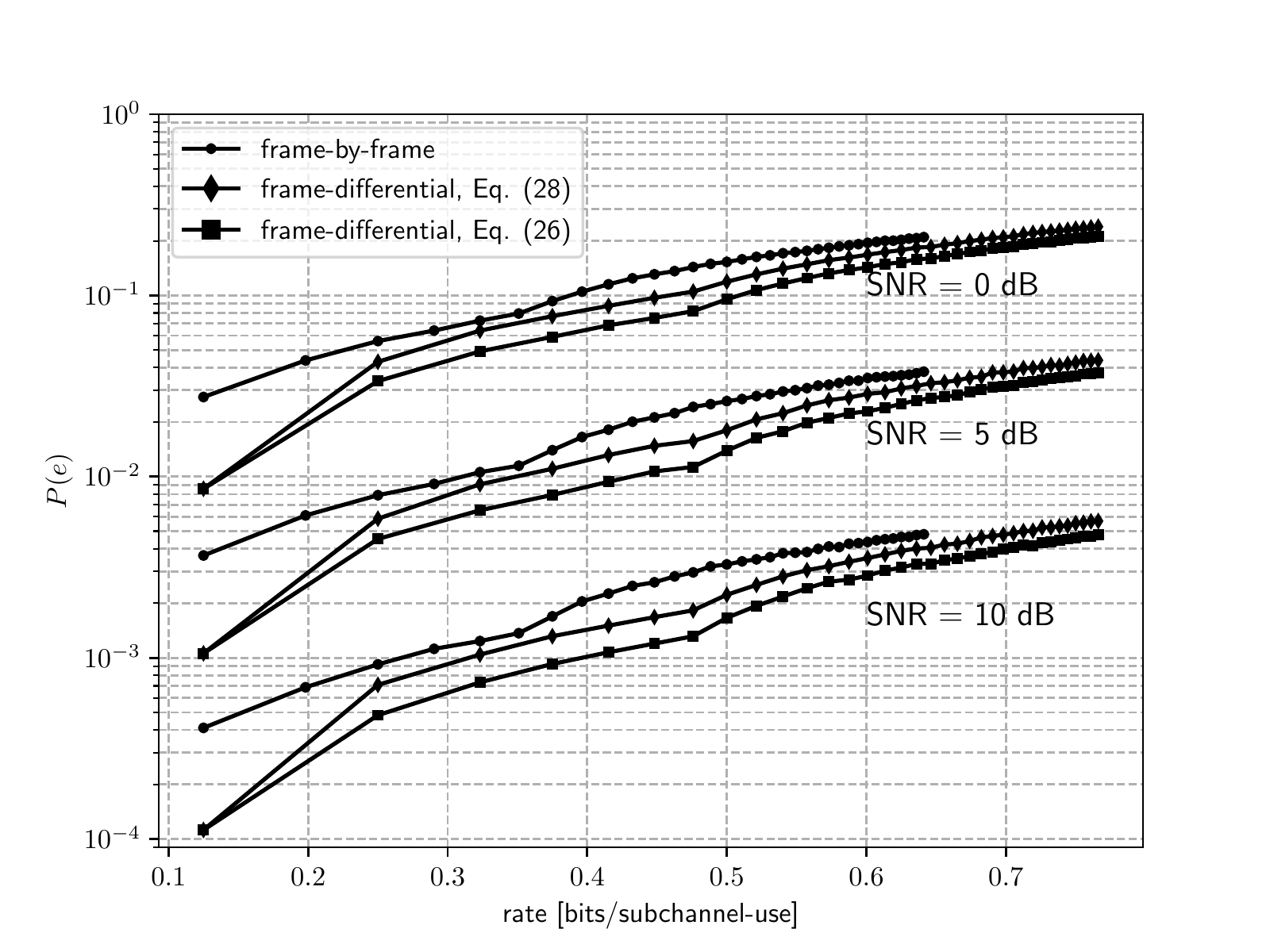}}
 \vspace{-0.2cm}
 \caption{$P(e)$ versus the transmission rate for $\text{SNR}=0,5,10$~dB, when $L=8$, $K=2$, $\rho=0$, $\kappa_{\alpha}=1/9$. For frame-differential encoding, $M=2$ and $|\mathcal{U}|$ is varied from $1$ to $35$; for frame-by-frame encoding, $|\mathcal{U}|$ is varied from $2$ to $35$.  \label{fig:differential-b}}

 \centerline{\includegraphics[width=\columnwidth]{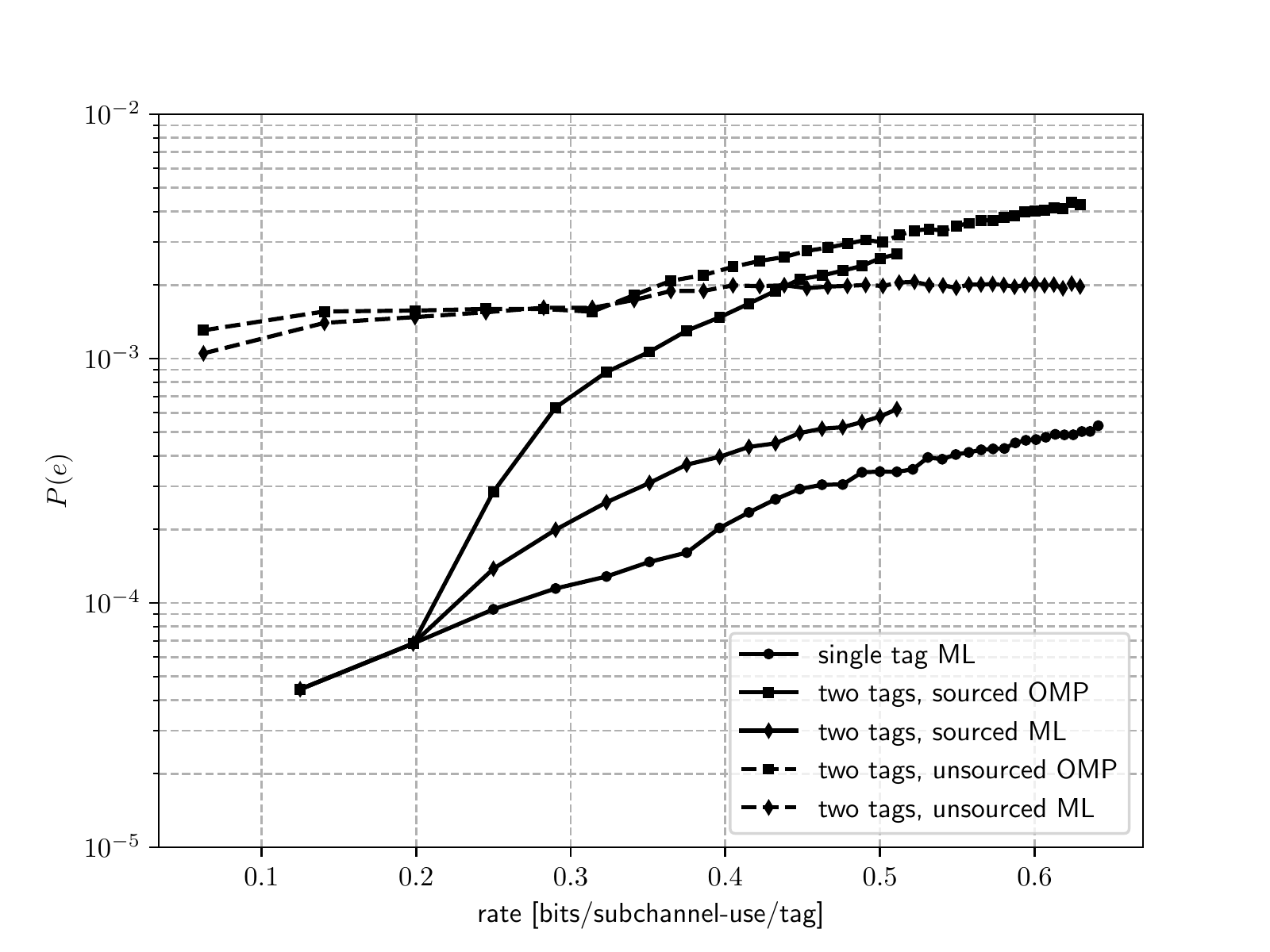}}
 \vspace{-0.2cm}
 \caption{$P(e)$ (averaged over all tags) versus the transmission  rate of each tag, when $1$ or $2$ tags are present,  $\text{SNR}=15$~dB, $L=8$, $K=2$, $\rho=0$, and $\kappa_{\alpha}=1/9$. Both a sourced and an unsourced multiple access is considered with ML or OMP-based decoding, as discussed in Sec.~\ref{SEC:Multiple-tag}.  \label{fig:multiuser}}
 \end{figure}

Next, we compare frame-differential and frame-by-frame encoding, when the same frame length and binary alphabet are employed. Fig.~\ref{fig:differential-a} shows $P(e)$  versus SNR when $D/L$ bits per subchannel-use are sent by the tag, for $D=1,3,5$: this is obtained with $M=2$ and $|\mathcal{U}|=2^{D-1}$ for frame-differential encoding and $|\mathcal{U}|=2^D$ for frame-by-frame encoding.
Also, Fig.~\ref{fig:differential-b} shows $P(e)$  versus the transmission rate, for $\text{SNR}=0,5,10$ dB. For frame-differential encoding, we only consider the decoding rules in~\eqref{joint_search} and~\eqref{rule_combined_3}. It is verified that frame-differential encoding provides a lower error probability for the same transmission rate; also, it can sustain larger transmission rates. This result indicates that the rate-splitting across the codebook $\mathcal{U}$ and the differential-encoded phase offset is advantageous, as it reduces the similarity among the messages. More generally, notice that the available degrees of freedom to get a desired transmission rate are the number of frames $L$, the size $M$ of the PSK constellation adopted for differential-encoding, and the codebook $\mathcal{U}$, which in principle may be jointly optimized to obtain a desired link quality and alphabet: this challenging problem is left for future studies. Finally,  notice in Fig.~\ref{fig:differential-b} that  $P(e)$ presents an evident slope variation after the $6$-th and $7$-th marker (left-to-right) for frame-by-frame and frame-differential encoding, respectively: this is a consequence of the fact that, beyond this point, it not possible to construct a codebook containing only orthogonal codewords (a similar effect is also present in Figs.~\ref{fig:cardU} and~\ref{fig:multiuser}).

\subsection{Multiple tags}

Finally, we consider the scenario discussed in Sec.~\ref{SEC:Multiple-tag}, when two active tags are present and the entries of all codewords are taken from a $2$-PSK alphabet. Fig.~\ref{fig:multiuser} shows $P(e)$  (averaged over all tags) versus the transmission rate of each tag for both a sourced and an unsourced multiple access, when $\text{SNR}=15$~dB and either ML-based or OMP-based decoding is employed; for comparison, the performance obtained with a single tag is also included (see also Fig.~\ref{fig:cardU}). In~\eqref{ML_dec_rule_0_unsourced} and in Algorithm~\ref{alg:omp_us}, we have numerically found and then used the threshold values $\gamma$ and $\eta$ providing the best error rate performance, respectively.

For a sourced multiple access, the largest codebooks $\mathcal U^{(1)}$ and $\mathcal U^{(2)}$ satisfying~\ref{P1s} and~\ref{P2s} have $|\mathcal U^{(1)}|=|\mathcal U^{(2)}|=17$; hence,  the curve for $Q=2$ ends at $(\log_2 17)/8=0.5109$ bits/subchannel-use/tag. When the transmission rate is $1/L$ or $(\log_{2}3)/L$ bits/subchannel-use/tag, $P(e)$ is the same for $Q=1,2$, as mutually-orthogonal codewords are assigned to all tags. For larger transmission rates, this orthogonality condition cannot be ensured anymore, and multiple tags are accommodated at the price of increasing $P(e)$; in this operating regime, ML-based decoding is superior to OMP-based  decoding, as it can better handle the multi-tag interference.

For an unsourced multiple access, the largest shared codebook $\mathcal U$ satisfying~\ref{P1} and~\ref{P2u} has  $|\mathcal U|=35$, as in the single-tag case. This multiple access strategy suffers a larger error probability than the sourced counterpart, as both the number of distinct messages and their content must be jointly estimated; however, it supports larger transmission rates. ML-based and  OMP-based decoding provide here similar performance up to a transmission rate of about $0.31$ [bits/subchannel-use/tag], corresponding to the $6$-th marker (left-to-right): this is because up to this point the adopted codebook contains orthogonal codewords. The minor difference is due to the fact OMP is an approximation of the decision rule in~\eqref{ML_dec_rule_0_unsourced}, which avoids the joint search over multiple messages under hypothesis $\mathcal{H}_{2}$. Beyond this point, we cannot construct a codebook containing orthogonal codewords and the gap increases.

\section{Conclusions}\label{SEC:Conclusions}
In this work, we have put forward the idea that the radar clutter can be used as a carrier signal to enable ambient backscatter communications. We have presented the signal model describing this system architecture and highlighted the interplay among the main system parameters. Upon exploiting the periodic structure of the radar clutter over time scales shorter than the channel coherence time, we have derived encoding/decoding strategies which allow the reader to distinguish the message sent by one or multiple tags from the superimposed radar interference without requiring a coordination with the radar transmitter or knowledge of the radar waveform and of the radar-tag-reader and radar-reader channels. 

We are now considering the design of frequency-domain signaling schemes and the development of more sophisticated multiple access protocols to support massive machine-type communications. Also, future studies should account for the Doppler effect and the imperfect synchronism between tag and reader. Finally, we foresee that the use of reconfigurable intelligent surfaces to implement an information-bearing tag or control the radar-tag-reader channel may greatly expand the potentiality of this idea by enlarging coverage and limiting the signal leakage towards undesired directions; in this context, an interesting research direction is optimizing the position, orientation, and beampattern of the tag/reader antenna, based on some prior cognition as to the surrounding environment and the radar location.

\appendices

\section{Proof of~\eqref{ML_dec_rule_0} and~\eqref{ML_dec_rule_multi}}\label{ML_appendix_multi}

Here we derive the ML decoding rule in~\eqref{ML_dec_rule_multi} for the observation model in~\eqref{rx_signal_multi}: when $Q=1$, we have $\bm X^{(1:Q)}=\bm x$, $\bm A^{(1:Q)} = \bm \alpha$, and $\mathcal{U}^{(1:Q)}_{s}=\mathcal{U}$, so that the model in~\eqref{rx_signal_multi} reduces to that in~\eqref{rx_signal_noindex}, and~\eqref{ML_dec_rule_multi} simplifies to~\eqref{ML_dec_rule_0}. To simplify the notation, we drop the subscripts in~\eqref{rx_signal_multi} and write the received signal as $\bm Y= \bm X\bm A\transp +\bm{1}_{L}\bm{i}\transp +\bm \Omega$. Accordingly, the ML estimator of $\bm X$ is
\begin{equation}
 \hat{\bm X}= \argmin_{\bm X \in \mathcal U^{(1:Q)}_{s}} \min_{\bm A \in \mathbb C^{K\times Q},\, \bm i \in \mathbb C^L} \Vert \bm Y - \bm X \bm A\transp - \bm 1_L \bm i\transp \Vert^2_F \label{x_est}
\end{equation}
where $\mathcal{U}^{(1:Q)}_{s}$ is the set in~\eqref{set_UQ}. Notice now that
\begin{multline}
 \Vert \bm Y - \bm X \bm A\transp - \bm 1_L \bm i\transp \Vert^2_F  = \trace ( \bm Y \bm Y\herm ) + \trace ( \bm X \bm A\transp \bm A^* \bm X\herm ) \\ + L\Vert \bm i \Vert^2 
 - 2\Re\{ \bm X\herm \bm Y \bm A^* + \bm 1_L\transp \bm Y \bm i^* - \bm 1_L\transp \bm X \bm A\transp \bm i^*\}
\end{multline}
and the conditions for the minimum over $(\bm \alpha, \bm i)$ are
\begin{subequations}
 \begin{align}
 L \bm i & = \bm Y\transp \bm 1_L - \bm A \bm X\transp \bm 1_L \label{i_der}\\
 \bm A \bm X\transp \bm X^* & = \bm Y\transp \bm X^* - \bm i \bm 1_L\transp \bm X^*.\label{alpha_der}
 \end{align}
\end{subequations}
Eliminating $\bm i$ in~\eqref{alpha_der}, we get
\begin{equation}
 \hat{\bm A} (\bm X) = \left( (\bm X\herm \bm P \bm X)^\dagger \bm X\herm \bm P \bm Y\right)\transp \label{alpha_est} 
\end{equation}
where $\bm P$ is the orthogonal projector in~\eqref{orth_proj_P}, and, plugging~\eqref{alpha_est} in~\eqref{i_der}, we obtain
\begin{equation}
 \hat{\bm i} (\bm X) = \frac{1}{L} \left( \bm 1\transp \left( \bm I_L - \bm X (\bm X\herm \bm P \bm X)^\dagger \bm X\herm \bm P \right) \bm Y\right)\transp . \label{i_est} 
\end{equation}
Finally, we have
\begin{align}
 \hat{\bm X} & = \argmin_{\bm X \in \mathcal \mathcal{U}^{(1:Q)}_{s}} \Vert \bm Y - \bm X \hat{\bm A}\transp (\bm X) - \bm 1_L \hat{\bm i}\transp (\bm X)\Vert^2_F \notag\\
 & = \argmin_{\bm X \in \mathcal \mathcal{U}^{(1:Q)}_{s}} \left\Vert \bm P \bm Y - \bm P \bm X(\bm X\herm \bm P \bm X)^\dagger \bm X\herm \bm P \bm Y \right\Vert^2_F \notag\\
 & = \argmin_{\bm X \in \mathcal \mathcal{U}^{(1:Q)}_{s}} \bigl\Vert \bigl( \bm I_L- \bm P \bm X(\bm X\herm \bm P \bm P\bm X)^\dagger \bm X\herm \bm P\bigr) \bm P \bm Y \bigr\Vert^2_F \notag\\
 & = \argmin_{\bm X \in \mathcal \mathcal{U}^{(1:Q)}_{s}} \bigl\Vert \bigl( \bm I_L- (\bm P \bm X)(\bm P \bm X)^\dagger \bigr) \bm P \bm Y \bigr\Vert^2_F \notag\\
 & = \argmax_{\bm X \in \mathcal \mathcal{U}^{(1:Q)}_{s}} \Vert \bm P \bm X (\bm P \bm X)^\dagger \bm P \bm Y \Vert^2_F \label{ML_rule_appendix}
\end{align}
where, in the last two equality, we have exploited the fact that $\bm B (\bm B\herm \bm B)^\dagger \bm B\herm = \bm B \bm B^\dagger$ is the orthogonal projector onto the range of a matrix $\bm B$, and that $\bm I - \bm B \bm B^\dagger$ is the orthogonal projector onto the null space of $\bm B\herm$. When $\bm X = \bm x$ is a column vector, we have
\begin{multline}
 \Vert \bm P \bm x (\bm P \bm x)^\dagger \bm P \bm Y \Vert^2_F  = \Vert \bm P \bm x (\bm x\herm \bm P \bm x)^{-1} \bm x\herm \bm P \bm Y \Vert^2_F \\ = \frac{\Vert \bm P \bm x \Vert^2 \Vert \bm x\herm \bm P \bm Y \Vert^2}{\Vert \bm P \bm x \Vert^4} = \frac{\Vert \bm x\herm \bm P \bm Y \Vert^2}{\Vert \bm P \bm x \Vert^2}
\end{multline}
and~\eqref{ML_rule_appendix} simplifies to~\eqref{ML_dec_rule_0}.

\end{document}